\documentclass[12pt,preprint]{aastex}

\newcommand{\ms}{\mbox{m s$^{-1}~$}}

\newcommand{\kms}{\mbox{km s$^{-1}~$}}
\newcommand{\msun}{M$_{\odot}$}

\newcommand{\mjup}{M$_{\rm JUP}$}

\newcommand{\msat}{M$_{\rm SAT}$}
\newcommand{\rjup}{R$_{\rm JUP}~$}
\newcommand{\msini}{$M \sin i~$}
\newcommand{\vsini}{$V \sin i~$}
\newcommand{\sini}{$\sin i~$}
\shortauthors{Vogt {\it et~al.\/}}
\shorttitle{Ten Low Mass Companions}
\slugcomment{WARNING: ROUGH DRAFT, Submitted to ApJ}
\received{}
\accepted{}
\begin{document}

\title{Ten Low Mass Companions from the \\
       Keck Precision Velocity Survey$~^{1}$}

\author{Steven S. Vogt\altaffilmark{2}, 
R. Paul Butler\altaffilmark{3}, 
Geoffrey W. Marcy\altaffilmark{4},
Debra A. Fischer\altaffilmark{4},
Dimitri Pourbaix\altaffilmark{5},
Kevin Apps\altaffilmark{6},
Gregory Laughlin\altaffilmark{2}}

\authoremail{vogt@ucolick.org}

\altaffiltext{1}{Based on observations obtained at 
Lick Observatory, which is operated by the University of
California, and on observations obtained at the
W.M. Keck Observatory, which is operated jointly by the
University of California and the California Institute of
Technology.  Keck time has been granted by both NASA and
the University of California.}

\altaffiltext{2}{UCO/Lick Observatory, 
University of California at Santa Cruz, Santa Cruz, CA, USA 95064}

\altaffiltext{3}{Department of Terrestrial Magnetism, Carnegie Institution
of Washington, 5241 Broad Branch Road NW, Washington D.C. USA 20015-1305}

\altaffiltext{4}{Department of Astronomy, University of California,
Berkeley, CA USA  94720 and at Department of Physics and Astronomy,
San Francisco State University, San Francisco, CA, USA 94132}

\altaffiltext{5}{F.N.R.S Post-doctoral researcher,
Institute of Astronomy and Astrophysics,
Universite Libre de Bruxelles CP 226 B-1050 Bruxelles, Belgium} 

\altaffiltext{6}{Physics and Astronomy, University of Sussex,
Falmer, Brighton BN1 9QJ, UK}

\begin{abstract}

Ten new low mass companions have emerged from the Keck precision
Doppler velocity survey, with minimum (\msini) masses ranging from
0.8 \mjup \ to 0.34 \msun.  Five of these are planet candidates
with \msini $<$ 12 \mjup, two are brown dwarf candidates with
\msini $\sim$30 \mjup, and three are low mass stellar companions.
Hipparcos astrometry reveals the orbital inclinations and masses 
for three of the (more massive) companions, and it provides 
upper limits to the masses for the rest.  A new class of extrasolar
planet is emerging, characterized by nearly circular orbits and 
orbital radii greater than 1 AU.  The planet HD 4208b appears to be
a member of this new class.  The mass distribution of extrasolar
planets continues to exhibit a rapid rise from 10 \mjup toward
the lowest detectable masses near 1 \msat.

\end{abstract}


\section{Introduction}
\label{intro}

All $\sim$70 known extrasolar planets have been discovered by the
precision Doppler technique over the last six years (Marcy, Cochran,
Mayor 2000).  These include four published multiple planet systems
(Butler et al. 1999; Marcy et al. 2001a, 2001b; Fischer et al. 2002)
and one transiting planet (Henry et al. 2000; Charbonneau et
al. 2000).  The mass function of giant planets is emerging (Marcy \&
Butler 2000; Jorissen, Mayor \& Udry 2001) and their occurrence
correlates with metallicity (Butler et al. 2000; Santos, Israelian, \&
Mayor 2001).

Precision Doppler surveys are biased toward finding massive companions
in small orbits, as these two characteristics enhance the Doppler
amplitude and allow many orbits to be observed quickly.  At the
current epoch, all of the confirmed extrasolar planets orbit within 4
AU of their host stars.  This is primarily due to the time baseline of
the active surveys, but also because the Doppler amplitude of an
orbiting companion decreases with increasing orbital radii.  A 1
\mjup \ planet in a 0.05 AU orbit induces a maximum Doppler amplitude
of 127 \ms on its star, but at 3 AU this is reduced to 16 \ms.  While
Doppler surveys with precision of 10 to 20 \ms can easily detect 1
\mjup \ planets at 0.05 AU, they are hard pressed to detect similar
planets at 3 AU.

Due to their larger mass, brown dwarf companions are much easier to
detect than planets.  A 20 \mjup \ brown dwarf companion at 0.05 AU
induces a maximum Doppler amplitude of 2500 \ms, a signal that
has been in the detectable range for most of the last century.
Yet no companion having mass between 10 and 80 \mjup \ has ever been
found orbiting within 0.1 AU of a star.

This paper reports the discovery of five planet candidates, two
brown dwarf candidates, and three low mass stellar companions from
the Keck precision velocity survey.  Hipparcos astrometric data is
used to solve for, or set limits on, the inclination angle and   
maximum mass of these companions.  Section 2 describes the Keck precision
velocity program.  The stellar properties and Keplerian orbital fits for
the ten new low mass companions are presented in Section 3.  
A discussion follows, including an updated substellar companion
mass function.

\section{The Keck Planet Search Program}
\label{obs}

The Keck Planet Search Program makes use of the HIRES echelle
spectrometer (Vogt et al. 1994) on the Keck I telescope.
The resolution of these spectra is R $\sim$ 80000, spanning
wavelengths from 3900--6200 \AA.  Wavelength calibration is
carried out by means of an iodine absorption cell
(Marcy \& Butler 1992) which superimposes a reference iodine
spectrum directly on the stellar spectra (Butler et al. 1996;
Valenti et al. 1995).  This system currently achieves
photon--limited measurement precision of 3 \ms (Vogt et al. 2000).
Efforts are under way to improve the single--shot precision
of this system to the 2 \ms level.

Observations from the Lick Observatory precision velocity
program are combined with the Keck observations for two of 
the stars presented here.  Lick observations are made with
either the 3--m Shane or 0.6--m Coud\'e Auxilliary (CAT)
Telescopes, both of which feed the ``Hamilton'' echelle
spectrometer (Vogt 1987).  Wavelength calibration at Lick is
also acomplished with an iodine absorption cell.

The Keck Planet Survey began in July 1996, and is currently
surveying about 650 main sequence dwarfs ranging in spectral
type from late F to mid--M.  The spectrum of stars earlier than
F7 do not contain enough Doppler information to achieve precision
of 3 \ms, while stars later than M5 are too faint even for Keck.

Nearly all suitable northern hemisphere G dwarfs within 50 pc,
and K dwarfs within 30 pc, are included in either the Keck survey
or the Lick 3--m survey (Fischer et al. 2002).  Evolved stars
have been removed from the observing list based on Hipparcos
distances (Perryman et al. 1997, ESA 1997).  The list has been further
sieved to remove chromospherically active stars as these stars
show velocity ``jitter'' of 10 to 50 \ms, related to rapid
rotation, spots, and magnetic fields (Saar et al. 1998).
The Ca II H\&K lines are used as a chromospheric diagnostic
(Noyes et al. 1984).  We measure the strength of the H\&K line
reversal directly from our Keck HIRES spectra.  Keck H\&K
measurements are placed on the Mt Wilson ``S'' scale by calibration
with previously published results (Duncan et al. 1991;
Baliunas et al. 1995; Henry et al. 1996).

Stars with known stellar companions within 2 arcsec are
removed from the observing list as it is operationally difficult
to get an uncontaminated spectrum of a star with a nearby
companion.  Otherwise there is no bias against observing
multiple stars.  The Keck target stars also contain no bias
against brown dwarf companions.

\section{New Companions from the Keck Survey}
\label{obs}

Five planet--mass candidates, 2 brown dwarf candidates, and 
3 stellar companions have emerged from the Keck survey.
The stellar properties of the 10 host stars are given
in Table 1.  The first two columns provide the HD catalog number
and  the Hipparcos catalog number respectively.  Spectral types are
from a calibration of $B-V$ and Hipparcos derived absolute magnitudes.
The stellar masses are estimated by interpolation of evolutionary
tracks (Fuhrmann 1998, Fuhrmann et al. 1997).  The [Fe/H] values
are drawn from a variety of sources (given in Section 3), 
including spectral synthesis matched directly
to our Keck HIRES spectra.

The R'$_{\rm HK}$ values, a chromospheric activity indicator
(Noyes et al. 1984), are measured from the CaII H\&K line cores
in our Keck spectra.  The level of Doppler ``jitter'' is correlated
with R'$_{\rm HK}$ (Saar et al. 1998).  Slowly rotating, chromospherically
inactive stars are intrinsically stable to at least the 3 \ms level,
while the Doppler ``jitter'' for young rapidly rotating stars ranges
from 10 to 50 \ms.

Figure 1 shows the H line for the 7 G dwarfs reported in this paper.
The Sun is shown for comparison.  Of these stars, HD 33636 is the
most active, with R'$_{\rm HK}$ $=$ $-4.81$, indicating a rotation period
of 13.6 d, and expected Doppler ``jitter'' of $\sim$7 \ms.  The other six G
dwarfs have R'$_{\rm HK}$ values similar or lower than the Sun, indicating
rotation periods of 25 days or longer, and intrinsic Doppler ``jitter'' of
3 \ms or less.

The H lines for the three K dwarfs reported in this paper are shown in 
Figure 2.  All three of these stars are slow rotators.  The rapid rotator
HD 128311 (K0 V) is shown at the bottome for comparison.  While even the
slowly rotating K dwarfs show a slight line reversal in the core of the H line,
they are clearly distinguished from rapid rotators like HD 128311.

The orbital parameters of the ten low mass companions are listed
in Table 2, while the individual Keck Doppler velocity measurements
are listed in Tables 3 through 12.  The host stars are discussed below.

The Hipparcos Intermediate Astrometric Data (ESA 1997) have been analyzed
with the orbits derived from the precision velocity data, using the
technique outlined by Pourbaix \& Arenou (2001) to constrain, or in 
three cases to solve for, the orbital inclination.

\subsection{HD 4208}

Based on Stromgren photometry, Eggen (1998) finds the metallicity of
HD 4208 (G5 V) to be [Fe/H] = -0.21, in good agreeement with our
estimate of -0.24 from spectral synthesis matched to our Keck HIRES
spectra. The star is photometrically stable at the level of Hipparcos
measurement error, $\sim$0.01 mag.  Based on the $B-V$ color, the
Hipparcos derived absolute magnitude, and the metallicity, we estimate
the mass of the primary to be 0.93 \msun.  This star is slowly
rotating and chromospherically inactive as indicated by the R'$_{\rm
HK}$ value.

Thirty--five Keck Doppler velocity observations have been made of HD 4208,
spanning 4.9 years, as shown in Figure 3 and listed in Table 3.  These
observations cover two full orbital periods.  The semiamplitude
($K$) of the Keplerian orbital fit is 18 \ms, only the third extrasolar
planet yet published with an amplitude less than 20 \ms (Marcy et al. 2000;
Fischer et al. 2002).  The RMS of the velocity residuals to the
Keplerian fit is 5.21 \ms, slightly worse than the typical internal measurement
error of 4.2 \ms.  The reduced $\chi_{\nu}^2$ of the Keplerian fit is 1.33.
Within measurement error the orbit is circular, thus joining
47 UMa (Butler \& Marcy 1996; Fischer et al. 2002) and HD 27442
(Butler et al. 2001) as the only published planets in circular
orbits beyond 0.2 AU.

With a semimajor axis $a$ = 1.67 AU, the maximum angular
separation between planet and star for HD 4208 is 53 milli-arcsec.
The orbit is similar to that of Mars in the Solar System.
The companion is not detected in the Hipparcos astrometric data,
thus constraining the orbital inclination to be larger than 1.4 deg.

Unlike most of the extrasolar planet candidates, HD 4208 is modestly
metal poor relative to the Sun.

\subsection{HD 114783}

We estimate the metallicity of HD 114783 (K2 V) to be [Fe/H] = +0.33
based on uvby photometry. Strassmeier et al. (2000) report that the
star is chromospherically inactive, in agreement with our
Keck--derived value of log(R'$_{\rm HK}$) = -4.96.  HD 114783 is
photometrically stable at the level of Hipparcos measurement error,
$\sim$0.01 mag.

A total of 37 Keck observations have been obtained between 1998 June and
2001 August, as shown in Figure 4 and Table 4.  These observations cover
slightly more than two full orbital periods of 501 days.  The semiamplitude
($K$) of the Keplerian orbital fit is 27 \ms, with an eccentricity of 0.10.
The RMS to the Keplerian fit is 4.08 \ms, with a reduced $\chi_{\nu}^2$ of 
1.48.  There is 5$\sigma$ discrepant point near 2000.35.  A periodogram
of the Keplerian residuals reveal no other periodicities with a false alarm
probability under 5\%.

At a distance of 20.4 pc, HD 114783 is an attractive astrometric target.
With a semimajor axis of 1.2 AU, the maximum separation between the planet
and the host star is 65 milli-arcsec.  The astrometric amplitude of
the star due to the planet is 61 mas/\sini, which should be easily
within the capabilities of both ground and space--based interferometry
within a few years.  The orbital inclination is constrained to be greater
than 1.26 degrees by the Hipparcos astrometric data.

\subsection{HD 4203}

HD 4203 (G5 V) is photometrically stable at the level of Hipparcos
measurement error, and chromospherically quiet with a measured
R'$_{\rm HK}$ value of -5.13.  At 77.8 pc, this is one of the
most distant stars on this project.  This star was added to
the Keck project in 2000 July based on the suggestion of
Laughlin (2000), who noted the star is metal rich ([Fe/H] = +0.22).  
>From interpolation of isochrones, Prieto \& Lambert (1999) estimate
the mass of the HD 4203 to be 0.98 \msun.  This estimate
does not take into account the extreme metallicity of
the star.  Based on its similarity to 51 Peg (Marcy et al. 1997),
we estimate the mass of HD 4203 to be 1.06 \msun.

A total of 14 Keck observations have been obtained between
2000 July and 2001 October.  These observations are listed in
Table 5 and graphically displayed in Figure 5.  The best--fit
Keplerian orbit to these data has a period of 406 d, a semiamplitude (K)
of 51 \ms, and an eccentricity of 0.53.  The minimum (\msini) mass
of the companion is 1.6 \mjup.  The RMS to the Keplerian fit
is 3.97 \ms, consistent with measurement uncertainty.

As only one orbit has been observed, the orbital period and
amplitude remain somewhat uncertain.  Only a few observations
have been obtained near the velocity maximum.  The next passage
through maximum velocity will occur in Spring 2002.  The orbital
inclination is constrained to be greater than 0.44 deg 
by Hipparcos astrometry.

\subsection{HD 68988}

Based on Stromgren photometry, Laughlin (2000) finds the metallicity
of HD 68988 (G2 V) to be [Fe/H] $=$ $+0.36$, consistent with the estimate
of Feltzing \& Gustafsson (1998), but somewhat higher than the +0.24
derived from matching spectral synthesis to our Keck HIRES spectra.
HD 68988 is photometrically stable at the level of Hipparcos measurement
error, $\sim$0.01 mag.  We estimate the mass of the primary to be
1.2 \msun. This star is slowly rotating and chromospherically inactive as
indicated by our measurement of the CaII line cores, 
log(R'$_{\rm HK}$) = -5.07.  The age of the star is estimated
to be approximately 6 Gyr based on the chromospheric diagnostic. 

Thirteen Keck Doppler observations and six Lick observations have been
made of HD 68988 spanning 1.4 years, as shown in Figure 6 and listed in
Table 6.  The orbital period is 6.276 d, the semiamplitude ($K$) is
187 \ms, and the eccentricity is $e$ $=$ $0.14$.  The RMS of the velocity
measurements to the best--fit Keplerian is 4.36 \ms, yielding a reduced
$\chi_{\nu}^2$ of 1.05.  The minimum mass of this companion is 1.9 \mjup.
In addition, there is a linear trend of -0.072 \ms per day, indicating a
second companion with an orbital period much greater than 4 years.

All of the published planets with orbital periods of less than 1 week\footnote{
HD 46375 (Marcy et al. 2000), HD 179949 (Tinney et al. 2001), HD 187123 (Butler
et al. 1998), $\tau$ Boo (Butler et al. 1997), BD--103166 (Butler et al. 2000),
HD 75289 (Udry et al. 2000, Butler et al. 2001), HD 209458 (Henry et al. 2000),
51 Peg (Mayor \& Queloz 1995; Marcy et al. 1997), $\upsilon$ And b (Butler et al. 1997;
Butler et al. 1999)} have circular orbits.  It is therefore surprising that
the orbit of HD 68988 is markedly non-circular.  This system is similar to the
companion orbiting HD 217107 (Fischer et al. 1999) with its period of 7.12 d,
\msini $=$ $1.25$ \mjup, and eccentricity of 0.134.  HD 68988 and
HD 217107 are both markedly metal rich. 

The time scale for orbital circularization due to tidal dissipation within
a planet is given by Marcy et al. (1997):
\begin{equation}
t_{circ}\approx {4 Q\over 63}{m\over M}{P_{orb}\over 2\pi}\left({a\over R_p}\right)^5.
\end{equation}
Assuming a radius 1.3 \rjup for the companion to HD 68988, this reduces to 5000Q/\sini years,
where Q is the  tidal quality factor (Goldreich \& Soter 1966).  For Q values less
than $\sim$ 10$^6$/\sini, the orbit of HD 68988 should have circularized.  The Q value
for Jupiter is inferred to be, 10$^5$ $<$ Q $<$ 10$^6$  (Goldreich \& Soter 1966;
Yoder 1979; Yoder \& Peale 1981).  The Q value for solid planets is in the range of
100 to 1000 (Marcy et al. 1997).  The planet orbiting HD 68988 is thus likely to be
a gas giant.  Three possibilities exist to explain the observed orbital eccentricity:
the planet has a higher Q value than is inferred for Jupiter; it
has only recently arrived in its current orbit; or it is being perturbed by
another body. 

\subsection{HD 33636}

Spectral synthesis matched to our Keck spectra yields [Fe/H] = -0.13
for HD 33636 (G0 V).  The star is photometrically stable at the level
of Hipparcos measurement error, and mildly active with 
log(R'$_{\rm HK}$) = -4.81.  The Doppler velocity ``jitter'' associated
with this level of activity for a G0 V star is $\sim$7 \ms
(Saar et al. 1998).  The Hipparcos derived distance of HD 33636 is 28.7 pc.  

A total of 21 observations of HD 33636 have been made at Keck
between 1998 January and 2001 October.  A further 11 observations
were made at Lick with the 3-m Shane telescope and the 0.6-m CAT
between 1998 Jan and 2001 Feb.  These observations are listed
in Table 7 and graphically displayed in Figure 7.  The best--fit
Keplerian orbit to the combined Keck and Lick data sets has a period
of 1553 d, a semiamplitude (K) of 148 \ms, and an eccentricity
of 0.39.  The minimum (\msini) mass of the companion is 7.7 \mjup.
The RMS to the Keplerian fit is 8.69 \ms, consistent with
the expected Doppler ``jitter'' due to activity.

The Keck and Lick observations of HD 33636 cover nearly the
full amplitude of the velocity variation, but slightly less than
one orbital period, leading to a large uncertainty in the derived
orbital period.  Hipparcos astrometry is not able to place limits
on the orbital inclination of this system because the orbital period
is much longer than the duration of the Hipparcos mission.

\subsection{HD 169822}

Spectral synthesis matched to our Keck spectra yield
[Fe/H] = -0.10 for HD 169822 (G5 V).  The star is chromospherically
quiet with R'$_{\rm HK}$ $=$ $-4.97$, and photometrically
stable at the level of Hipparcos measurement error.

We have obtained a total of 22 Keck observations between 1999 July
and 2001 October.  The measured Doppler velocities are shown
in Figure 8 and listed in Table 8.  The best--fit Keplerian
to these velocities yields a period of 293 d, an eccentricity
of 0.48, and a semiamplitude ($K$) of 991 \ms.  The minimum
(\msini) mass of the companion is 27.2 \mjup.  The RMS to
the Keplerian is 5.70 \ms.

The Hipparcos catalog assigned a stochastic solution
for HD 169822 because none of the investigated models
gave a satisfactory solution at that time.  The
Hipparcos data has been re--investigated in light
of the Doppler velocity signal, and now clearly
yields a periodicity of 292.7 days, in excellent
agreement with the Doppler velocity data.  The new
Hipparcos solution, taking into account the orbital
motion, changes the measured parallax to this system
from 37.04 mas to 31.2 mas, increasing the distance
of this system from 27 to 32 pc.  The Hipparcos--derived 
orbital inclination of 175 degrees yields
a companion mass of 0.30 \msun.  The companion is
thus stellar, presumably with a spectral type around
M3 V.  With a semimajor axis of 0.84 AU, the maximum
separation between the companion and the primary
is 39 milli-arcsec.  The M dwarf companion is expected
to be $\sim$5.5 $V$ magnitudes fainter than the
primary.

\subsection{HD 184860}

The metallicity of HD 184860 (K2 V), determined by
spectral synthesis matched to our Keck spectra,
is [Fe/H] $=$ $-0.13$.  The star is chromospherically
quiet, with R'$_{\rm HK}$ $=$ $-5.00$.  Hipparcos $V$
band photometric scatter is 0.016 mag.

A total of 21 Keck Doppler velocity of HD 184860 have
been made between 1996 July and 2001 August.
These observations are shown graphically in Figure 9
and in tabular form in Table 9.  The best--fit Keplerian
yields an orbital period of 693.0 d, a semiamplitude
$K$ $=$ 1123 \ms, eccentricity $e$ $=$ 0.67, and
\msini $=$ $32$ \mjup.  As no observations have been made
near minimum velocity, the semiamplitude is poorly
constrained.  Minimum velocity will next occur in 
2002 December.

With a semi--major axis of 1.44 AU, the companion  to
HD 184860 is separated from the primary by 47.5 milli-arsec.
The minimum astrometric semiamplitude of the primary
is $\sim$1.8 milli-arsec.  Unfortunately, HD 184860 has a stellar
companion separated by 5.1 arcsec.  This stellar companion
shows up in the Hipparcos Intermediate Astrometric Data,
making it impossible to search for the astrometric signal
that is caused by the brown dwarf candidate.  It is thus not possible
to derive an upper limit for the orbital inclination
of the brown dwarf candidate.  Moderate values of \sini
would render this an M dwarf.

\subsection{HD 64468}

Spectral synthesis matched to our Keck spectra yield
[Fe/H] $=$ $+0.00$ for HD 64468 (K2 V).  The star is chromospherically
quiet with R'$_{\rm HK}$ $=$ $-5.03$, and photometrically stable
within Hipparcos measurement error.

Thirteen observations of HD 64468 have been made at Keck
between 1997 Jan and 2001 April.  These observations are listed
in Table 10 and graphically displayed in Figure 10.  The best--fit
Keplerian orbit to this data has a period of 161 d, a semiamplitude ($K$)
of 5730 \ms, and an eccentricity of 0.26.  The minimum (\msini) mass
of the companion is 0.13 \msun.  As the companion is clearly stellar,
we will drop this star from our observing program.

The orbital inclination of this system was derived
from the Hipparcos astrometric data by freezing the
orbital parameters from the Doppler velocities, 
yielding an inclination of 64 $\pm$13 degrees, and 
a true mass for the companion of 0.14 \msun.  
The companion is likely to be an M6 dwarf.
With a semimajor axis of 0.56 AU, the maximum
separation between the companion and the primary
is 28 milli-arcsec.  The M dwarf companion is
expected to be $\sim$7.5 $V$ magnitudes fainter
than the primary.

\subsection{HD 35956A}

Based on Stromgren photometry, Eggen (1998) reports the metallicity
of HD 35956A (G0 V) to be [Fe/H] = -0.14, in reasonable agreeement
with our photometric estimate of -0.21.  The star is photometrically
stable at the level of Hipparcos measurement error, $\sim$0.01 mag,
and chromospherically quiet with R'$_{\rm HK}$ = -4.92.
The Hipparcos derived distance of this star is 28.9 pc.  
We estimate the mass of the primary to be 1.0 \msun.

We have obtained a total of 14 Keck observations between 1996 October
and 2001 October.  The measured Doppler velocities are shown
in Figure 11 and listed in Table 11.  The best--fit Keplerian
orbit to these velocities yields a period of 1427 d, an eccentricity
$e$ $=$ $0.62$, and a semiamplitude, $K$ = 3796 \ms.  The RMS
to the Keplerian fit is 4.89 \ms.  The minimum (\msini) mass of
the companion is 0.18 \msun.  As this companion is stellar, we
will drop this star from our target list.

Although the orbital period of this binary is about 50\% longer
than the duration of the Hipparcos mission, the astrometric
signature of the companion was detected.  Freezing the orbital
elements from the precision velocities yields an orbital inclination
of 78 $\pm$4 degrees.  The true mass of the companion is thus
only 2\% larger than the minimum mass.  The spectral type of
the companion is expected to be roughly M6, with $V$ magnitude
$\sim$16, separated from the $V$ $=$ $6.7$ primary by 0.1 arcsec. 

\subsection{HD 43587}

HD 43587 is assigned a spectral type of F9 V by Simbad and G0.5 by
Hipparcos.  The $B-V$ color of $0.61$ is consistent with G1 or G2
spectral type.  Spectral synthesis matched to our Keck spectra yields
Teff = 5795 K, [Fe/H] = -0.03, and \vsini = 2.7 \kms.  The R'$_{\rm
HK}$ value measured from the H \& K lines in the Keck spectra is
-4.97.  The star is photometrically stable at the level of Hipparcos
measurement error.  HD 43587 is thus a near solar twin.

Fourteen observations of HD 43587 have been made at Keck between 1996
Oct and 2001 April.  These observations are listed in Table 12 and
graphically displayed in Figure 12.  The best--fit Keplerian to this
data has a period of 33.7 yr, a semiamplitude (K) of 4323 \ms, and an
eccentricity of 0.80.  The minimum (\msini) mass of the companion is
0.34 \msun.  Given the long orbital period, we were fortunate to
observe the extreme velocity drop of 1998.  As the companion is
clearly stellar, we will drop this star from our observing program.
Hipparcos was unable to astrometrically detect the companion due to
the short duration of the Hipparcos mission.

This star was observed 7 times during 5.3 yr around epoch 1987 by the
CORAVEL Doppler velocity survey (Duquennoy \& Mayor 1991), and found
to be constant, with standard deviation of 0.36 \kms.  That constant
velocity is consistent with our results.  The present best--fit
Keplerian predicts the last epoch of rapid velocity change would have
been in 1964.  During the epoch when CORAVEL measurements were made,
the velocity of the star would have changed by only 0.2 \kms per year.
We expect that in retrospect, the CORAVEL velocities should reveal a
trend just above the errors.  The absolute velocity reported by
Duquennoy \& Mayor was 9.6 \kms. We find that on 2 Dec 1997 (JD =
2450785) the velocity of HD43587 was 12.3$\pm$0.1 \kms (Nidever et
al. 2002).  Thus, the measured change in the velocity of the star from
1987 to 2 Dec 1997 was +2.7 \kms . In comparison, our orbital solution
predicts a change of +2.4 \kms during that 10--year period, in good
agreement within errors.

With a semimajor axis a = 11.6 AU, the companion orbits about
0.6 arcseconds from the primary.  For a minimum mass companion
(\sini = 1), the primary would be about 5.5 V magnitudes
brighter than the companion.  The value of \sini is likely
greater than 0.5, as smaller values would yield a companion
with an expected difference in V magnitude of 3 or less,
which would be detectable in our Keck spectra.  This binary
thus makes an interesting AO target.  

\section{Discussion}

A total of 1,200 stars are currently being surveyed by the
Keck, Lick, and Anglo--Australian precision velocity surveys.
All three of these programs use the Iodine cell technique,
and all three have demonstrated long term precision of 3 \ms
(Butler \& Marcy 1997; Vogt et al. 2000; Butler et al. 2001).
These three surveys have no selection bias against stars with
brown dwarf companions.

A total of 48 substellar candidates (\msini $<$ 80 \mjup)
have been uncovered from these surveys, of which 44 are 
planet candidates with plausible masses below the Deuterium
burning limit (\msini $<$ $12$ \mjup).
All of these companions have either been published or are
currently submitted to referee journals, including 3 new
companions from the AAT survey (Tinney et al. 2002;
Jones et al. 2002).

Two of the four brown dwarf candidates from these surveys,
HD 169822 and HD 164427 (Tinney et al. 2001), have been
revealed by Hipparcos astrometry to be M dwarfs.  Hipparcos
is able to place upper limits on the masses of the remaining
49 companions (Pourbaix \& Arenou 2001; Perryman et al. 1996).

The mass function of the 46 surviving substellar candidates from
the Keck, Lick, and Anglo--Australian surveys is shown in
Figure 13.  All but 3 of these companions orbit within 3 AU.
The mass function is flat and sparsely populated above 10 \mjup,
and then begins rising abruptly below 10 \mjup.

Within 3 AU these surveys are complete for companions having
\msini $>$ 10 \mjup.  Incompleteness
is greatest for the smallest mass bin, \msini $<$ 1 \mjup.
Companions with \msini $<$ 1 \msat \ are only detectable if
they orbit within 1 AU.  These selection effects strongly
favor the detection of companions with \msini $>$ 10 \mjup \ at
the expense of companions with \msini $<$ 1 \mjup.  In spite of
this, 15 companions having \msini $<$ 1 \mjup \ have been
detected, while only two have been found with \msini $>$ 10 \mjup.

The abrupt and discontinuous jump in the mass function below
10 \mjup \ empirically sets the threshold
between planets and brown dwarfs at $\sim$10 \mjup.  Coincidentally
the Deuterium-burning limit resides near this limit at 12 \mjup.

The two companions with the largest mass in Figure 13 are HD 168443c
(\msini = 17 \mjup) and HD 184860 (\msini = 32 \mjup).  As described
in the previous section, Hipparcos can not put an upper limit on the
mass of the companion to HD 184860 because of a nearby stellar companion.
A moderate inclination of 23 degrees would be be sufficient to push this
brown dwarf candidate into the M dwarf mass range.  For the case of
HD 168443c, Hipparcos astrometry is able to put an upper limit on the
mass of the companion of 42 \mjup \ (Marcy et al. 2001a).  
Brown dwarf companions orbiting within 3 AU of main sequence stars are
rare, with at most about 1 such companion per 500 stars.  In contrast, 
planetary mass companions are common.

Relatively few planets have been found orbiting metal poor stars
(Butler et al. 2000, Santos et al. 2001).  Planets orbiting stars of
less than solar metallicity ([Fe/H] $<$ $+0.00$) are indicated in
Figure 13 by the cross hatched area.  They constitute about 12\% of
the planets found in the Keck, Lick, and Anglo-Australian surveys
(even though two--thirds of the field stars have metallicity less than
solar).  If planets are common around metal poor stars, these planets
must either have smaller masses or larger orbital radii than the
planets found to date.

\acknowledgements

We acknowledge support by NSF grant AST-9988358 (to SSV), NSF grant
AST-9988087 and travel support from the Carnegie Institution of
Washington (to RPB), NASA grant NAG5-8299 and NSF grant AST95-20443
(to GWM), and by Sun Microsystems.  We thank the NASA and UC Telescope
assignment committees for allocations of telescope time.  This
research has made use of the Simbad database, operated at CDS,
Strasbourg, France.

\clearpage
\begin{figure}
\caption{Ca II H line cores for seven G dwarfs in ascending
order of $B-V$.  The HD catalog number of each star is
shown along the right edge.  The Sun is shown for comparison.
With the exception of HD 33636, the R'$_{\rm HK}$ values 
derived from the H\&K lines are similar to the Sun, indicating
rotation periods of 25 d or longer and photospheric Doppler
``jitter'' of 3 \ms or less.  For HD 33636 the derived
rotation period is 13 d, and the expected Doppler ``jitter''
is $\sim$7 \ms.}
\label{H_Gdwarfs}
\end{figure}

\begin{figure}
\caption{Ca II H line cores for three K dwarfs in ascending 
order of $B-V$.  The HD catalog number of each star is
shown along the right edge.  The active K0 dwarf HD 128311
is shown for comparison.  Even slowly rotating K dwarfs show mild
line core reversal.  Dramatic line core reversal is seen in the
rapidly rotating, chromospherically active, K0 V star HD 128311.}
\label{H_Kdwarfs}
\end{figure}

\begin{figure}
\caption{Doppler velocities for HD 4208 (G5 V).
The solid line is a Keplerian orbital fit with a
period of 829 days, a semiamplitude of 18.3 \ms,
and an eccentricity of 0.04, yielding a minimum
(\msini) of 0.80 \mjup \ for the companion.  The
RMS of the Keplerian fit is 5.21 \ms.}

\label{Kdwarfs}
\end{figure}

\begin{figure}
\caption{Doppler velocities for HD 114783 (K2 V).
The solid line is a Keplerian orbital fit with a
period of 501 days, a semiamplitude of 27 \ms,
and an eccentricity of 0.10, yielding a minimum
(\msini) of 1.0 \mjup \ for the companion.  The
RMS of the Keplerian fit is 4.08 \ms.}
\label{fig4}
\end{figure}

\begin{figure}
\caption{Doppler velocities for HD 4203 (G5 V).
The solid line is a Keplerian orbital fit with a
period of 406 days, a semiamplitude of 51 \ms,
and an eccentricity of 0.53, yielding
\msini $=$ 1.6 \mjup \ for the companion.  The
RMS of the Keplerian fit is 3.97 \ms.}
\label{fig5}
\end{figure}

\begin{figure}
\caption{Phased Doppler velocities for HD 68988 (G2 V).
The filled dots are the Keck 10--m observations, while
the Lick 3--m observations are indicated by the crosses.
The solid line is the best--fit Keplerian orbit to the
combined data sets.  The period is 6.276 d, the semiamplitude 
is 187 \ms, and the eccentricity is 0.14, yielding
(\msini) $=$ 1.9 \mjup \ for the companion.  The
RMS of the Keplerian fit is 4.36 \ms.
A linear trend of -26.4 /ms per year has been
removed.  This linear trend suggests a second companion
with an orbital period much longer than 4 years.}
\label{fig13}
\end{figure}

\begin{figure}
\caption{Doppler velocities for HD 33636 (G0 V).
The filled dots are the Keck 10--m observations, while
the Lick 3--m observations are indicated by the crosses.
The solid line is a Keplerian orbital fit with a
period of 4759 d, a semiamplitude of 188 \ms,
and an eccentricity of 0.67, yielding a minimum
(\msini) of 11.5 \mjup \ for the companion.  The
RMS of the Keplerian fit is 8.75 \ms.  Nearly
the full amplitude of the velocity variation has
been observed, but only a faction of the total
orbit.  Large uncertainties therefore remain in
the orbital period and minimum mass of the
companion.}
\label{fig7}
\end{figure}

\begin{figure}
\caption{Doppler velocities for HD 169822 (G5 V).
The solid line is a Keplerian orbital fit with a
period of 293 d, a semiamplitude of 991 \ms,
and an eccentricity of 0.48, yielding a minimum
(\msini) of 27.2 \mjup \ for the companion.  The
RMS of the Keplerian fit is 5.92 \ms.  Hipparcos
astrometry finds the inclination i $=$ 175 degrees,
and the true mass of the companion to be 0.30 \msun.
The companion is therefore an M dwarf in a nearly
face--on orbit.}
\label{fig8}
\end{figure}

\begin{figure}
\caption{Doppler velocities for HD 184860 (K2 V).
The solid line is a Keplerian orbital fit with a
period of 693 d, a semiamplitude of 1123 \ms,
and an eccentricity of 0.67, yielding a minimum
(\msini) of  32 \mjup \ for the companion.  The
RMS of the Keplerian fit is 7.82 \ms.  Hipparcos
astrometry is unable to constrain \sini due
to a known nearby stellar companion.  This companion
may well be an M dwarf in a modestly inclined orbit.}
\label{fig9}
\end{figure}

\begin{figure}
\caption{Doppler velocities for HD 64468 (K2 V).
The solid line is a Keplerian orbital fit with a
period of 161 d, a semiamplitude of 5730 \ms,
and an eccentricity of 0.26, yielding a minimum
(\msini) of 0.13 \msun \ for the companion.  The
RMS of the Keplerian fit is 16.2 \ms.  The Hipparcos
derived orbital inclination is 64 degrees, yielding
a true mass of 0.14 \msun.  The companion is thus
an M dwarf.}
\label{fig10}
\end{figure}

\begin{figure}
\caption{Doppler velocities for HD 35956 A (G0 V).
The solid line is a Keplerian orbital fit with a
period of 1427 d, a semiamplitude of 3796 \ms,
an eccentricity of 0.62, and  (\msini) $=$ 0.18 \msun 
for the companion.  The RMS of the Keplerian fit is
4.86 \ms.  The Hipparcos derived orbital inclination
is 78 degrees, thus the true mass is only 2\% larger
than the minimum (\msini) mass. The companion is 
an M dwarf.}
\label{fig11}
\end{figure}

\begin{figure}
\caption{Doppler velocities for HD 43587 (G2 V).
The solid line is a Keplerian orbital fit with a
period of 33.7 yr, a semiamplitude of 4323 \ms,
and an eccentricity of 0.80, yielding a minimum
(\msini) of 0.34 \msun \ for the companion.  The
RMS of the Keplerian fit is 8.83 \ms.  While only
10\% of the orbital period has been covered, the
full amplitude was observed between 1997 and 1999.
Hipparcos astrometry is unable to detect binaries
with such long orbital periods.}
\label{fig12}
\end{figure}

\begin{figure}
\caption{Substellar mass function found from the Keck, Lick,
and AAT precision Doppler surveys.  These are the only surveys
sensitive to 1 \mjup \ planets orbiting beyond 3 AU.  Out to 3 AU
these surveys are complete for companions of more than 5 \mjup.
Incompleteness is greatest for the smallest mass bin.  The
discontinuous and abrupt rise in the mass function below 10 \mjup \
empirically motivates setting the upper mass limit of planets
near 10 \mjup.  Planets orbiting stars with [Fe/H] $<$ $+0.0$ are
indicated by by cross hatching.  About one--third of field stars
are metal rich relative to the Sun, but roughly 88\% of the planets
found from these surveys orbit metal rich stars.}
\label{fig13}
\end{figure}

\clearpage

\begin{deluxetable}{rrrllllll}
\tablenum{1}
\tablecaption{Stellar Properties}
\label{stellar}
\tablewidth{0pt}
\tablehead{
Star & Star & Spec   & M$_{\rm Star}$ & V     & B-V & log(R'$_{\rm HK}$) & [Fe/H] & d   \\
(HD) & (Hipp) & type & (M$_{\odot}$)  & (mag) &     &                    &        & (pc) 
} 
\startdata
  4203 &   3502 & G5 ~V & 1.06 & 8.70 & 0.771 & -5.13 & +0.22 & 77.8 \\
  4208 &   3479 & G5 ~V & 0.93 & 7.78 & 0.664 & -4.93 & -0.24 & 32.7 \\
 33636 &  24205 & G0 ~V & 0.99 & 7.00 & 0.588 & -4.81 & -0.13 & 28.7 \\
35956A &  25662 & G0 ~V & 1.00 & 6.71 & 0.582 & -4.92 & -0.21 & 28.9 \\
 43587 &  29860 & G2 ~V & 1.02 & 5.70 & 0.610 & -4.97 & -0.03 & 19.3 \\
 64468 &  38657 & K2 ~V & 0.78 & 7.76 & 0.950 & -5.03 & +0.00 & 20.0 \\
 68988 &  40687 & G2 ~V & 1.20 & 8.20 & 0.652 & -5.07 & +0.24 & 58.8 \\
 114783 & 64457 & K2 ~V & 0.92 & 7.56 & 0.930 & -4.96 & +0.33 & 20.4 \\
 169822 & 90355 & G5 ~V & 0.91 & 7.83 & 0.699 & -4.97 & -0.10 & 32.0 \\
 184860 & 96471 & K2 ~V & 0.80 & 8.38 & 1.011 & -5.00 & -0.13 & 30.3 \\
\enddata
\end{deluxetable}

\clearpage

\begin{deluxetable}{rrllllllll}
\tablenum{2}
\rotate
\tablecaption{Orbital Parameters}
\label{candid}
\tablewidth{0pt}
\tablehead{
\colhead{Star}  & \colhead{Period} & \colhead{$K$} & \colhead{$e$} & \colhead{$\omega$} & \colhead{$T_0$} & \colhead{\msini} & \colhead{a} & \colhead{N} & \colhead{RMS} 
\\
\colhead{(HD)} & \colhead{(days)} & \colhead{(\ms)} &\colhead{ } & \colhead{(deg)} & \colhead{(JD-2450000)}  & \colhead{(\mjup)} & \colhead{(AU)} & \colhead{obs } & \colhead{(\ms)}
} 
\startdata
4208  &  829 (36)  & 18.3 (2)  & 0.04 (0.12) & 301 (84) & 1774 (197) & 0.80 & 1.7 & 35 & 5.21 \\
114783 &  501 (14)  & 27   (2)  & 0.10 (0.08) &  97 (40) & 1840 (59)  & 1.0  & 1.2 & 37 & 4.08 \\
4203  &  406 (30)  & 51   (5)  & 0.53 (0.10) & 271 (50) & 1882 (5) & 1.6 & 1.1 & 14 & 3.97 \\
68988\tablenotemark{a} & 6.276 (0.002) & 187 (6) & 0.14 (0.03) & 186 (8) & 1913.0 (0.2) & 1.9 & 0.071 & 19 & 4.36 \\
33636 &  1553  (800) & 148  (15)  & 0.39 (0.09) & 335 (8) & 1196 (21) & 7.7 & 2.6 & 32 & 8.69 \\
169822&  293.1 (0.5)  & 991 (87) & 0.48 (0.03) & 173 (1) & 1919 (1) & 27.2 & 0.84 & 22 & 5.70 \\
184860 &  693 (1) & 1123 (490) & 0.67 (0.06) & 132 (6) & 1906 (9) & 32.0 & 1.4 & 21 & 7.82 \\
64468 & 161.2 (0.1) & 5730 (7) & 0.262 (0.002) & 328.1 (0.2) & 457.0 (0.2) & 139 & 0.56 & 13 & 16.2 \\
35956A &  1427 (1) & 3796 (4) & 0.616 (0.002) & 326.5 (0.2) & 796.3 (0.3) & 184 & 2.6 & 14 & 4.86 \\
43587  &  12325 (500) & 4323 (9) & 0.80 (0.01) & 75 (1) & 0832 (1) & 358 & 11.6 & 14 & 8.83 \\
\enddata
\tablenotetext{a}{Additional Velocity Slope is -26.4 $\pm$5.5 \ms per yr.}
\end{deluxetable}

\clearpage

\begin{deluxetable}{rrr}
\tablenum{3}
\tablecaption{Velocities for HD 4208}
\label{vel4208}
\tablewidth{0pt}
\tablehead{
JD & RV & error \\
(-2450000)   &  (m s$^{-1}$) & (m s$^{-1}$)
}
\startdata
   366.9657  &     8.5  &  3.5 \\
   715.0257  &   -17.4  &  3.6 \\
   786.7220  &   -19.6  &  5.6 \\
  1010.1052  &    13.3  &  3.1 \\
  1014.0991  &    17.6  &  3.9 \\
  1043.0497  &    13.4  &  3.7 \\
  1044.0338  &     9.4  &  3.2 \\
  1068.9389  &     9.0  &  3.5 \\
  1172.7428  &     9.0  &  4.2 \\
  1368.0796  &   -13.4  &  3.4 \\
  1412.0645  &   -18.6  &  3.8 \\
  1438.9302  &   -26.6  &  4.0 \\
  1543.7389  &   -13.5  &  4.7 \\
  1550.7261  &   -10.8  &  4.2 \\
  1551.7132  &   -24.1  &  4.5 \\
  1552.7113  &   -11.8  &  4.2 \\
  1580.7057  &   -13.7  &  4.1 \\
  1581.7067  &   -29.2  &  4.2 \\
  1582.7244  &   -24.8  &  4.4 \\
  1583.7066  &   -20.8  &  4.2 \\
  1585.7075  &   -22.7  &  4.8 \\
  1755.0330  &     7.7  &  4.1 \\
  1756.0257  &     4.2  &  3.9 \\
  1757.0736  &     5.8  &  3.6 \\
  1793.9300  &     4.9  &  4.1 \\
  1882.7286  &    18.0  &  4.7 \\
  1883.7727  &    24.7  &  4.9 \\
  1899.7776  &    21.6  &  4.5 \\
  1900.7493  &    17.0  &  4.6 \\
  2095.1118  &    -4.6  &  4.5 \\
  2129.0961  &     4.2  &  5.0 \\
  2133.0468  &    -6.6  &  4.2 \\
  2134.0092  &    -4.5  &  4.5 \\
  2161.9285  &   -14.4  &  4.3 \\
  2187.9813  &    -8.7  &  4.7 \\
\enddata
\end{deluxetable}

\clearpage

\begin{deluxetable}{rrr}
\tablenum{4}
\tablecaption{Velocities for HD 114783}
\label{vel114783}
\tablewidth{0pt}
\tablehead{
JD & RV & error \\
(-2450000)   &  (m s$^{-1}$) & (m s$^{-1}$) 
}
\startdata
   983.7917  &   -22.9  &  2.8 \\
  1200.0942  &    29.1  &  2.8 \\
  1310.9285  &     5.6  &  2.8 \\
  1370.8121  &   -17.8  &  3.4 \\
  1551.1677  &   -11.6  &  2.8 \\
  1552.1549  &    -8.9  &  2.3 \\
  1581.1187  &     3.2  &  2.5 \\
  1582.0853  &    -0.2  &  2.5 \\
  1583.0644  &     2.9  &  2.8 \\
  1584.1043  &     2.1  &  2.5 \\
  1585.0266  &     2.8  &  1.7 \\
  1586.0245  &     4.8  &  2.7 \\
  1678.8802  &    21.9  &  2.8 \\
  1679.8972  &    10.8  &  2.6 \\
  1702.9153  &    28.5  &  2.6 \\
  1703.8098  &    29.9  &  2.5 \\
  1704.8819  &    26.5  &  2.8 \\
  1705.8815  &    28.8  &  2.7 \\
  1755.7586  &    28.7  &  3.3 \\
  1884.1633  &   -17.7  &  2.7 \\
  1898.1701  &   -25.5  &  2.4 \\
  1899.1755  &   -23.3  &  2.7 \\
  1900.1717  &   -18.6  &  2.3 \\
  1901.1800  &   -25.4  &  2.6 \\
  1972.1100  &   -22.4  &  2.5 \\
  1973.1352  &   -23.7  &  2.9 \\
  1974.1353  &   -26.1  &  2.8 \\
  1975.1449  &   -28.0  &  2.7 \\
  1982.1079  &   -21.9  &  3.0 \\
  2002.9808  &   -20.6  &  3.0 \\
  2003.8969  &   -16.5  &  3.0 \\
  2009.0445  &   -23.0  &  3.2 \\
  2030.9115  &   -13.8  &  3.1 \\
  2062.8241  &     8.2  &  4.0 \\
  2094.7722  &     9.3  &  3.1 \\
  2100.7892  &     1.4  &  3.0 \\
  2127.7826  &    19.7  &  3.4 \\
\enddata
\end{deluxetable}

\clearpage

\begin{deluxetable}{rrr}
\tablenum{5}
\tablecaption{Velocities for HD 4203}
\label{vel4230}
\tablewidth{0pt}
\tablehead{
JD & RV & error \\
(-2450000)   &  (m s$^{-1}$) & (m s$^{-1}$) 
}
\startdata
   757.1224  &   -26.2  &  3.6 \\
   792.9725  &   -32.4  &  3.8 \\
   882.8351  &     5.3  &  3.4 \\
   883.8483  &    11.6  &  3.9 \\
   900.8378  &    45.6  &  3.4 \\
  1063.1258  &     2.8  &  4.0 \\
  1065.1288  &    10.6  &  4.6 \\
  1096.1144  &    -7.7  &  3.7 \\
  1097.0676  &    -3.3  &  4.6 \\
  1128.1161  &    -5.0  &  3.8 \\
  1133.0558  &    -8.6  &  3.6 \\
  1133.9261  &   -19.5  &  3.6 \\
  1162.9185  &   -26.6  &  3.8 \\
  1187.9624  &   -32.6  &  3.6 \\
\enddata
\end{deluxetable}

\clearpage

\begin{deluxetable}{rrrr}
\tablenum{6}
\tablecaption{Velocities for HD 68988}
\label{vel68988}
\tablewidth{0pt}
\tablehead{
JD & RV & error & Keck\\
(-2450000)   &  (m s$^{-1}$) & (m s$^{-1}$) & Lick 
}
\startdata
   552.0229  &  -102.4  &  4.0 & K \\
   582.8602  &  -129.4  &  4.4 & K \\
   899.1245  &   126.6  &  3.5 & K \\
   901.1357  &    48.6  &  3.5 & K \\
   913.9219  &     6.2  &  8.0 & L \\
   914.8319  &  -139.6  &  8.8 & L \\
   915.7716  &  -150.0  &  7.5 & L \\
   927.7894  &  -146.3  &  8.8 & L \\
   945.8093  &   -99.5  &  8.5 & L \\
   946.7834  &  -157.2  & 10.8 & L \\
   972.0226  &  -159.7  &  3.9 & K \\
   972.9965  &   -87.9  &  4.3 & K \\
   973.8936  &    36.8  &  3.2 & K \\
   974.8776  &   193.9  &  3.5 & K \\
   982.9828  &    -6.5  &  3.8 & K \\
  1003.7900  &  -141.8  &  3.9 & K \\
  1007.8684  &    29.5  &  4.5 & K \\
  1062.7491  &   179.6  &  3.8 & K \\
  1064.7680  &   -57.4  &  3.8 & K \\
\enddata
\end{deluxetable}

\clearpage

\begin{deluxetable}{rrrr}
\tablenum{7}
\tablecaption{Velocities for HD 33636}
\label{vel33636}
\tablewidth{0pt}
\tablehead{
JD & RV & error & Keck\\
(-2450000)   &  (m s$^{-1}$) & (m s$^{-1}$) & Lick 
}
\startdata
   831.7357  &   -87.6  & 30.1 & L \\
   838.7594  &   -82.1  &  4.5 & K \\
  1051.1034  &    38.1  &  3.9 & K \\
  1073.0403  &    61.9  &  3.6 & K \\
  1154.7925  &   150.9  & 11.3 & L \\
  1171.8449  &   167.4  &  3.3 & K \\
  1228.8034  &   200.0  &  3.9 & K \\
  1412.1067  &   112.1  &  4.3 & K \\
  1447.0323  &   101.4  & 10.5 & L \\
  1543.8997  &    28.5  &  4.2 & K \\
  1550.8857  &    19.7  &  2.9 & K \\
  1580.8356  &    15.2  &  4.7 & K \\
  1581.8678  &    11.6  &  4.0 & K \\
  1582.7846  &    10.3  &  4.1 & K \\
  1607.6845  &     8.5  & 12.8 & L \\
  1628.6289  &    22.9  & 12.0 & L \\
  1793.1197  &   -38.1  &  4.6 & K \\
  1859.9447  &   -65.5  &  9.3 & L \\
  1860.9068  &   -64.6  & 10.2 & L \\
  1882.9336  &   -69.1  &  4.3 & K \\
  1884.0852  &   -67.0  &  4.0 & K \\
  1898.0322  &   -73.2  &  3.9 & K \\
  1899.0454  &   -71.6  &  3.6 & K \\
  1900.0648  &   -63.7  &  3.7 & K \\
  1901.0137  &   -60.9  &  3.4 & K \\
  1913.7822  &   -86.2  &  6.1 & L \\
  1914.8443  &   -87.5  &  7.9 & L \\
  1915.8011  &   -82.7  &  8.0 & L \\
  1945.7179  &   -80.6  &  5.7 & L \\
  1973.7486  &   -90.4  &  5.8 & K \\
  2003.7459  &   -83.0  &  4.1 & K \\
  2188.1390  &   -91.2  &  4.4 & K \\
\enddata
\end{deluxetable}

\clearpage

\begin{deluxetable}{rrr}
\tablenum{8}
\tablecaption{Velocities for HD 169822}
\label{vel169822}
\tablewidth{0pt}
\tablehead{
JD & RV & error \\
(-2450000)   &  (m s$^{-1}$) & (m s$^{-1}$) 
}
\startdata
   372.8809  &  -554.1  &  3.5 \\
   373.8011  &  -529.2  &  3.9 \\
   411.8501  &    14.6  &  3.0 \\
   679.0826  &  -292.6  &  3.3 \\
   680.1159  &  -274.4  &  2.8 \\
   703.0212  &   -10.2  &  2.7 \\
   703.9911  &     5.3  &  2.7 \\
   705.0419  &    13.3  &  4.4 \\
   705.9615  &    25.1  &  3.0 \\
   707.0839  &    28.1  &  4.8 \\
   755.9132  &   239.8  &  4.6 \\
   793.8101  &   291.8  &  3.8 \\
   973.1655  &  -283.7  &  4.5 \\
   982.1650  &  -150.0  &  3.6 \\
  1004.1285  &    64.0  &  3.3 \\
  1009.1085  &    93.0  &  3.4 \\
  1030.9853  &   210.0  &  3.6 \\
  1061.9457  &   293.2  &  3.9 \\
  1062.9627  &   278.2  &  3.8 \\
  1094.8889  &   276.3  &  4.4 \\
  1128.8559  &   174.4  &  3.8 \\
\enddata
\end{deluxetable}

\clearpage

\begin{deluxetable}{rrr}
\tablenum{9}
\tablecaption{Velocities for HD 184860}
\label{vel184860}
\tablewidth{0pt}
\tablehead{
JD & RV & error \\
(-2450000)   &  (m s$^{-1}$) & (m s$^{-1}$) 
}
\startdata
   283.9812  &   480.9  &  4.2 \\
   602.0837  &  -747.4  &  3.1 \\
   665.9255  &  -321.5  &  3.1 \\
   955.0729  &   441.1  &  2.8 \\
   955.9625  &   439.2  &  3.8 \\
   984.0518  &   466.1  &  6.4 \\
  1011.8909  &   526.0  &  7.9 \\
  1050.8529  &   576.5  &  5.0 \\
  1311.0711  &  -604.0  &  4.6 \\
  1367.8866  &  -284.3  &  4.8 \\
  1409.8883  &  -112.6  &  4.8 \\
  1438.7611  &   -26.9  &  5.0 \\
  1439.7771  &   -23.5  &  5.7 \\
  1679.0771  &   470.6  &  6.5 \\
  1703.0113  &   513.6  &  6.7 \\
  1793.8075  &   647.8  &  6.2 \\
  1975.1678  &  -876.1  &  4.3 \\
  1982.1679  &  -813.8  &  6.0 \\
  2004.1326  &  -606.1  &  4.5 \\
  2096.9296  &  -130.8  &  9.4 \\
  2128.8532  &   -31.8  &  5.4 \\
\enddata
\end{deluxetable}

\clearpage

\begin{deluxetable}{rrr}
\tablenum{10}
\tablecaption{Velocities for HD 64468}
\label{vel64468}
\tablewidth{0pt}
\tablehead{
JD & RV & error \\
(-2450000)   &  (m s$^{-1}$) & (m s$^{-1}$)
}
\startdata
   462.9003  &  7537.9  &  3.0 \\
   545.8154  & -3502.0  &  4.5 \\
   807.0220  &  4765.3  &  3.1 \\
   839.0313  &  -897.6  &  4.7 \\
   861.9171  & -3106.4  &  5.0 \\
  1170.9913  & -2030.9  &  4.5 \\
  1226.9085  & -2685.2  &  5.0 \\
  1550.9977  & -2413.7  &  4.8 \\
  1552.9947  & -2112.1  &  5.8 \\
  1581.9092  &  5895.1  &  4.7 \\
  1898.1031  &  3965.0  &  3.1 \\
  1973.8651  & -1665.3  &  6.0 \\
  2003.7708  & -3750.1  &  5.0 \\
\enddata
\end{deluxetable}

\clearpage

\begin{deluxetable}{rrr}
\tablenum{11}
\tablecaption{Velocities for HD 35956A}
\label{vel35956A}
\tablewidth{0pt}
\tablehead{
JD & RV & error \\
(-2450000)   &  (m s$^{-1}$) & (m s$^{-1}$) 
}
\startdata
   366.1155  & -2018.8  &  3.9 \\
   546.7296  & -1974.0  &  3.8 \\
   786.9244  &  4363.6  &  4.8 \\
   837.8768  &  5357.2  &  3.3 \\
  1051.1210  &   972.3  &  3.9 \\
  1170.9331  &   -16.7  &  3.9 \\
  1229.7433  &  -354.8  &  4.3 \\
  1543.9305  & -1533.8  &  4.6 \\
  1550.8895  & -1544.7  &  4.3 \\
  1551.8912  & -1540.9  &  4.2 \\
  1884.0873  & -2068.4  &  4.7 \\
  2003.7483  & -1879.5  &  4.8 \\
  2128.1349  &  -340.6  &  4.2 \\
  2188.1412  &  2578.7  &  3.5 \\
\enddata
\end{deluxetable}

\clearpage

\begin{deluxetable}{rrr}
\tablenum{12}
\tablecaption{Velocities for HD 43587}
\label{vel43587}
\tablewidth{0pt}
\tablehead{
JD & RV & error \\
(-2450000)   &  (m s$^{-1}$) & (m s$^{-1}$)
}
\startdata
   366.1316  &  5052.2  &  3.2 \\
   545.7812  &  5586.7  &  4.5 \\
   787.0277  &  3819.9  &  3.7 \\
   807.0639  &  3252.6  &  3.4 \\
   838.9376  &  2236.4  &  3.6 \\
   861.7638  &  1481.6  &  3.7 \\
  1069.0974  & -2481.5  &  4.0 \\
  1171.8867  & -2895.6  &  4.4 \\
  1227.8236  & -2943.0  &  5.1 \\
  1544.0042  & -2810.6  &  3.9 \\
  1552.9211  & -2788.6  &  3.9 \\
  1582.8475  & -2752.9  &  4.3 \\
  1884.0956  & -2446.7  &  4.5 \\
  2003.7636  & -2310.7  &  6.3 \\
\enddata
\end{deluxetable}

\clearpage


\clearpage

\begin{thebibliography}{}
\parsep 0pt
\itemsep -3pt

\bibitem[Baliunas et~al. 1995]{Bal95}
Baliunas, S.~L., Donahue, R.~A., Soon, W.~H., Horne, J.~H.,
Frazer, J., Woodard-Eklund, L., Bradford, M., Rao, L.~M.,
Wilson, O.~C., Zhang, Q., Bennett, W., Briggs, J.,
Carroll, S.~M., Duncan, D.~K., Figueroa, D., Lanning, H.~H.,
Misch, A., Mueller, J., Noyes, R.~W., Poppe, D., Porter, A.~C.,
Robinson, C.~R., Russell, J., Shelton, J.~C., Soyumer, T.
Vaughan, A.~H., \& Whitney, J.~H. 1995,
\newblock {  ApJ, } {438}, 269.

\bibitem[Butler { et~al.} 1996]{BuMaWi96} Butler, R.~P., Marcy, G.~W.,
Williams, E., McCarthy, C., Dosanjh, P., \& Vogt, S.~S. 1996,
\newblock { PASP, } {108}, 500

\bibitem[Butler \& Marcy 1996]{BuMa96}
Butler, R.~P. \&  Marcy, G.~W. 1996,
\newblock {  ApJ, } {464}, L153.

\bibitem[Butler {  et~al.} 1997]{BuMaWi97}
Butler, R.~P., Marcy, G.~W., Williams, E., Hauser, H., \& Shirts, P.  1997,
\newblock {  ApJ, } {474}, L115

\bibitem[Butler \& Marcy 1997]{BuMa97}
Butler, R.~P. \& Marcy, G.~W. 1997,
``The Near Term Future of Extrasolar Planet Searches'',
Brown Dwarfs and Extrasolar Planets, held on Tenerife, 17-21 March 1997,
ed. R. Rebolo, E.L. Martin, and M.R. Zapatero Osorio,
ASP Conference Series, Vol. 134, p. 162. 

\bibitem[Butler { et~al.} 1998]{BuMa98}
Butler, R.~P., Marcy, G.~W., Vogt, S.~S. \& Apps, K. 1998,
\newblock { PASP, } {110}, 1389.

\bibitem[Butler {  et~al.} 1999]{BuMaFi99}
Butler, R.~P., Marcy, G.~W., Fischer, D.~A., Brown, T.~M.,
Contos, A.~R., Korzennik, S.~G., Nisenson, P. \& Noyes, R.~W. 1999,
\newblock {  ApJ, } {526}, 916.

\bibitem[Butler { et~al.} 2000]{BuMa00}
Butler, R.~P., Vogt, S.~S., Marcy, G.~W., Fischer, D.~A.,
Henry, G.~W. \& Apps, K. 2000,
\newblock { ApJ, } {545}, 504.

\bibitem[Butler { et~al.} 2001]{BuMa01}
Butler, R.~P., Tinney, C.~G., Marcy, G.~W., Jones, H.~R.~A.,
Penny, A.~J.  \& Apps, K. 2001,
\newblock { ApJ, } {555}, 410.

\bibitem[Charbonneau { et~al.} 2000]{Char00}
Charbonneau, D., Brown, T.~M., Latham, D.~W. \& Mayor, M. 2000,
\newblock { ApJ, } {529}, L49.

\bibitem[Duquennoy \& Mayor 1991]{Duq91}
Duquennoy, A. \& Mayor, M. 1991,
\newblock {  A\&A, } {248}, 485.

\bibitem[Duncan { et~al.} 1991]{Dun91}
Duncan, D.~K., Vaughan, A.~H., Wilson, O.~C., Preston, G.~W., 
Frazer, J., Lanning, H.~H., Misch, A., Mueller, J., Soyumer, D.,
Woodard, L., Baliunas, S.~L., Noyes, R.~W., Hartmann, L.~W., 
Porter, A.~C., Zwaan, C., Middelkoop, F., Rutten, R.~G.~M.,
Mihalas, D. 1991,
\newblock {  ApJS, } {76}, 383.

\bibitem[Eggen 1998]{Eggen98}
Eggen, O.~J. 1998,
\newblock {  AJ, } {115}, 2397.

\bibitem[ESA 1997]{ESA97}
ESA 1997, The Hipparcos and Tycho Catalogues (ESA SP-1200).

\bibitem[Feltzing \& Gustafsson 1998]{FelGus98}
Feltzing, S. \&  Gustafsson, B. 1998,
\newblock {  A\&AS, } {129}, 237.

\bibitem[Fischer {  et~al.} 1999]{FiMa99}
Fischer, D.~A., Marcy, G.~W., Butler, R.P., Vogt, S.~S.\& Apps, K. 1999,
\newblock {  PASP, } {111}, 50 

\bibitem[Fischer {  et~al.} 2002]{Fisch02}
Fischer, D.~A., Marcy, G.~W., Butler, R.P., Laughlin, G. \& Vogt, S.~S. 2002,
\newblock {  ApJ, } to appear 10 Jan 2002.

\bibitem[Fuhrmann 1998]{Fuhr98}
Fuhrmann,K. 1998, {  A\&A }, 338, 161.

\bibitem[Furhmann {  et~al.} 1997]{Fuhr97}
Fuhrmann, K., Pfeiffer, M.J. \& Bernkopf, J. 1997,
\newblock {  A\&A, } {326}, 1081.

\bibitem[Goldreich & Soter 1966]{GoSo66}
Goldreich, P. \& Soter, S. 1966, 
\newblock {  Icarus, } {5}, 375.

\bibitem[Henry {  et~al.} 2000]{Henry00}
Henry, G.~W., Marcy, G.~W., Butler, R.~P. \& Vogt, S.~S. 2000,
\newblock { ApJ, } {529}, L45.

\bibitem[Henry { et~al.} 1996]{Henry96}
Henry, T.~J., Soderblom, D.~R., Donahue, R.~A. \& Baliunas, S.~L. 1996,
\newblock {  AJ, } {111}, 439

\bibitem[Jones et al. 2002]{jones:02}
Jones, H.~R.~A., Butler, R.~P., Tinney, C.~G., Marcy, G.~W.,
Penny, A.~J., McCarthy, C., Carter, B.~D. \& Apps, K. 2002,
\newblock {  ApJ, }, submitted.

\bibitem[Jorissen 2001]{jor01}
Jorissen, A.,  Mayor, M. \& Udry, S. 2002, 
\newblock {  A\&A, } accepted.

\bibitem[Laughlin 2000]{laugh00}
Laughlin, G. 2000,
\newblock {  ApJ, } {545}, 1064.

\bibitem[Marcy \& Butler 1992]{MaBu92}
Marcy, G.~W. \&  Butler, R.~P. 1992,
\newblock {  PASP, } {104}, 270.

\bibitem[Marcy \& Butler 2000]{MaBu00}
Marcy, G.~W. \&  Butler, R.~P. 2000,
\newblock {  PASP, } {112}, 137.

\bibitem[Marcy { et~al.} 2000]{MaBuVo00}
Marcy, G.~W., Butler, R.~P., Vogt, S.~S. 2000,
\newblock {  ApJ, } {536}, L43.

\bibitem[Marcy { et~al.} 1997]{MaBuWi97} Marcy, G.~W., Butler, R.~P.,
Williams, E., Bildsten, L., Graham, J.~R., Ghez, A., \& Jernigan, G.
1997, \newblock { ApJ, } 481, 926.

\bibitem[Marcy { et~al.} 2000]{MaCoMa00}
Marcy, G.~W., Cochran, W.~D. \& Mayor, M. 2000, 
in Protostars and Planets IV, ed. V. Mannings, A. P. Boss \&
S. S. Russell (Tucson: University of Arizona Press), p1285.

\bibitem[Marcy {  et~al.} 2001a]{mbv2001a}
Marcy, G.~W., Butler, R.~P., Vogt, S.~S., Liu, M.~C.,
Laughlin, G.~P., Apps, K., Graham, J.~R., Lloyd, J.,
Luhman, K.~L. \& Jaywardhana, R. 2001a,
\newblock { ApJ, } 555, 418.

\bibitem[Marcy {  et~al.} 2001b]{mbv2001b}
Marcy, G.~W., Butler, R.~P., Fischer, D.~A.,
Vogt, S.~S., Lissauer, J.~J. \& Rivera, E.~J. 2001b,
\newblock { ApJ, } 556, 296.

\bibitem[Mayor \& Queloz 1995]{Mayor95}
Mayor, M. \& Queloz, D. 1995, Nature,378,355

\bibitem[Nidever { et~al. } 2002]{Nide02}
Nidever, D., Marcy, G.~W., Butler, R.~P. \& Fischer, D.~A. 2002,
\newblock {  PASP, } to be submitted.

\bibitem[Noyes et al. 1984]{noyes:84} Noyes, R.~W., Hartmann, L.,
Baliunas, S.~L., Duncan, D.~K., \& Vaughan, A.~H. 1984,
\newblock {  ApJ, } 279, 763.

\bibitem[Perryman et al. 1996]{Perry96}
Perryman, M.~A.~C., et al. 1996, { A\&A, } 310, L21. 

\bibitem[Perryman et al. 1997]{Perry97}
Perryman, M.~A.~C., et al. 1997, { A\&A, } 323, L49. The Hipparcos Catalog

\bibitem[Pourbaix & Arenou 2001]{Pour2001}
Pourbaix, D. \& Arenou, F. 2001,
\newblock {  A\&A, } 372, 935.

\bibitem[Prieto & Lambert 1999]{Priet99}
Prieto, C.~A. \& Lambert, D.~L. 1999, { A\&A, } 352, 555.

\bibitem[Saar {  et~al.} 1998]{SaBuMa98}
Saar, S.~H., Butler, R.~P., \& Marcy, G.~W. 1998,
\newblock {  ApJ, } 403, L153.

\bibitem[Santos {  et~al.} 2001]{Santos01}
Santos, N. C., Israelian, G., Mayor, M.  2001, A\&A, 373, 1019

\bibitem[Strassmeier{  et~al.} 2000]{Strass00}
Strassmeier, K.~G., Washuettl, A., Granzer, T., Scheck,
M. \&  Weber, M. 2000,
\newblock {  A\&A Supp, }, 142, 275.

\bibitem[Tinney et al. 2001]{tinney:01}
Tinney, C.~G., Butler, R.~P., Marcy, G.~W., Jones, H.~R.~A.,
Penny, A.~J., Vogt, S.~S., Henry, G.~W. 2001,
\newblock {  ApJ, }, 551, 507.

\bibitem[Tinney et al. 2002]{tinney:02}
Tinney, C.~G., Butler, R.~P., Marcy, G.~W., Jones, H.~R.~A.,
Penny, A.~J., McCarthy, C. \& Carter, B.~D. 2002,
\newblock {  ApJ, }, submitted.

\bibitem[Udry et al. 2000]{udry:00}
Udry, S., Mayor, M., Naef, D., Pepe, F., Queloz, D., Santos, N.,
Burnet, M., Confino, B. \& Melo, C. 2000, { A\&A, } 356, 590.

\bibitem[Valenti { et~al.} 1995]{VaBuMa95} 
Valenti, J., Butler, R.~P. \& Marcy, G.~W. 1995
\newblock { PASP, } {107}, 966.

\bibitem[Vogt 1987]{vogt:87}
Vogt, S.~S. 1987, { PASP, } 99, 1214.

\bibitem[Vogt et al. 1994]{vogt:94}
Vogt, S.~S. et al. 1994, Proc. Soc. Photo-Opt. Instr. Eng., 2198, 362

\bibitem[Vogt {  et~al.} 2000]{vogt:00}
Vogt, S.~S., Marcy, G.~W., Butler, R.~P. \& Apps, K. 2000,
\newblock { ApJ, } {536}, 902.

\bibitem[Yoder 1979]{yoder:79}
Yoder, C. F. 1979, 
\newblock { Nature, } {279}, 767.

\bibitem[Yoder & Peale 1981]{yoder:81}
Yoder, C. F. \& Peale, S. J. 1981, 
\newblock { Icarus, } {47}, 1.

\end{thebibliography}
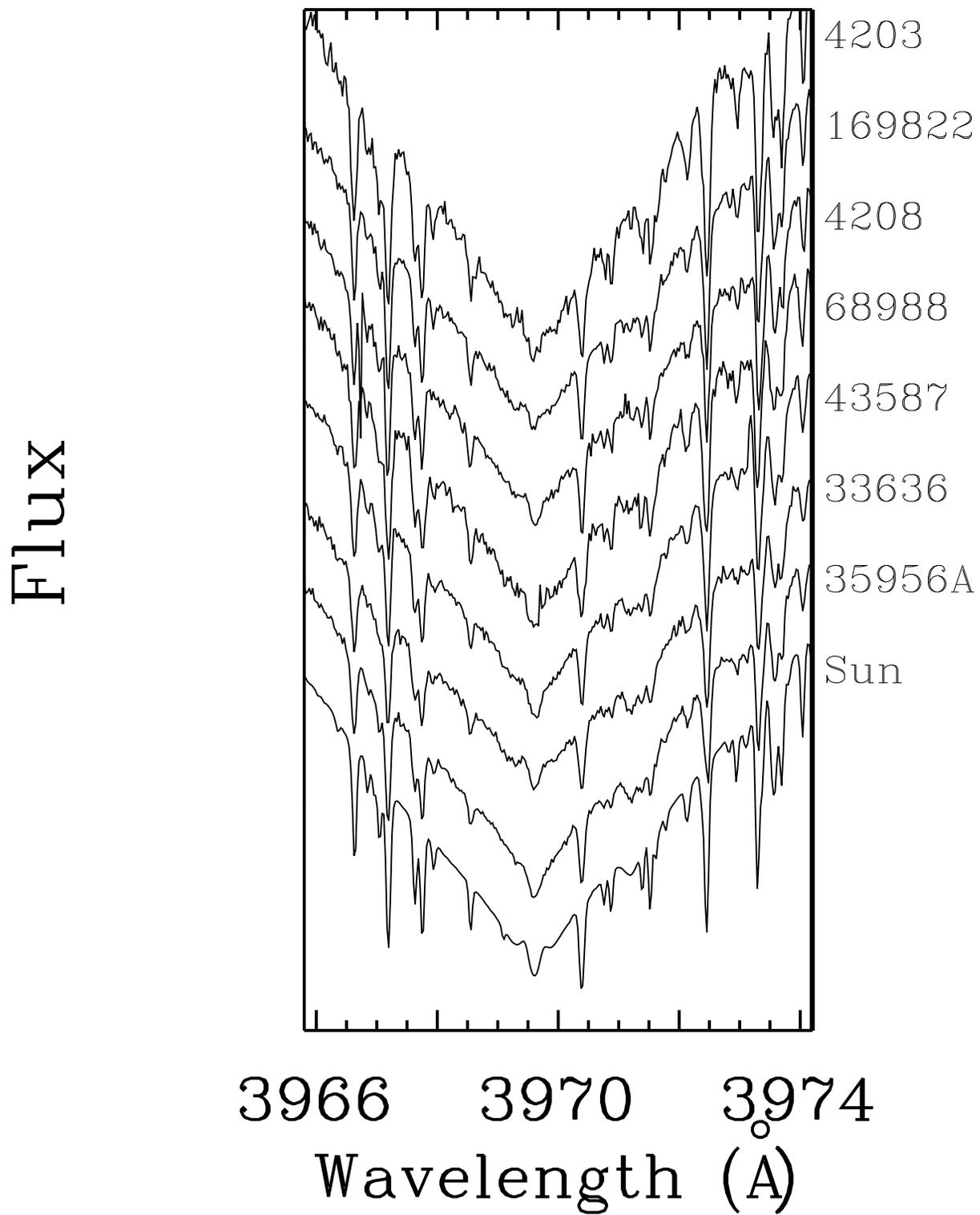
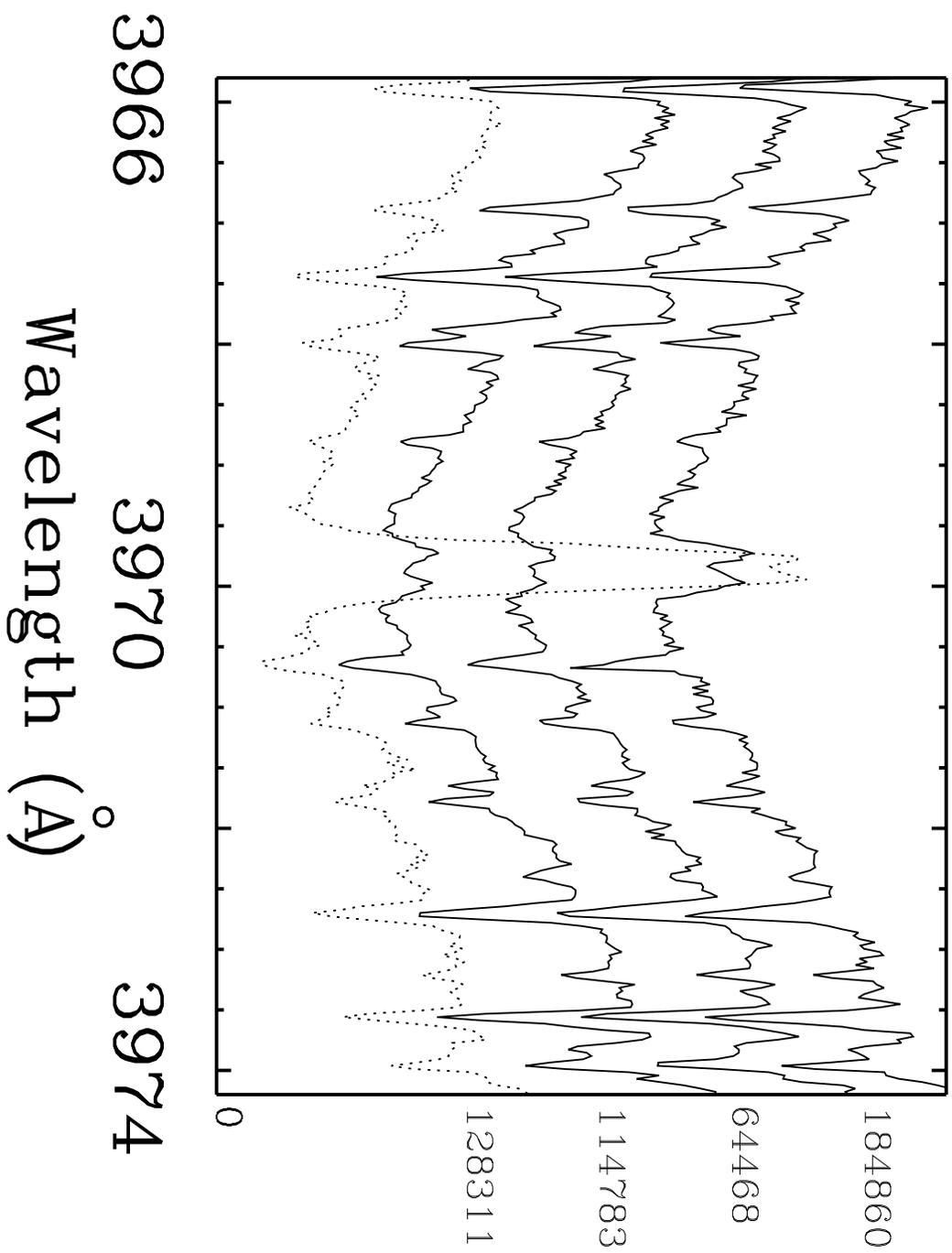
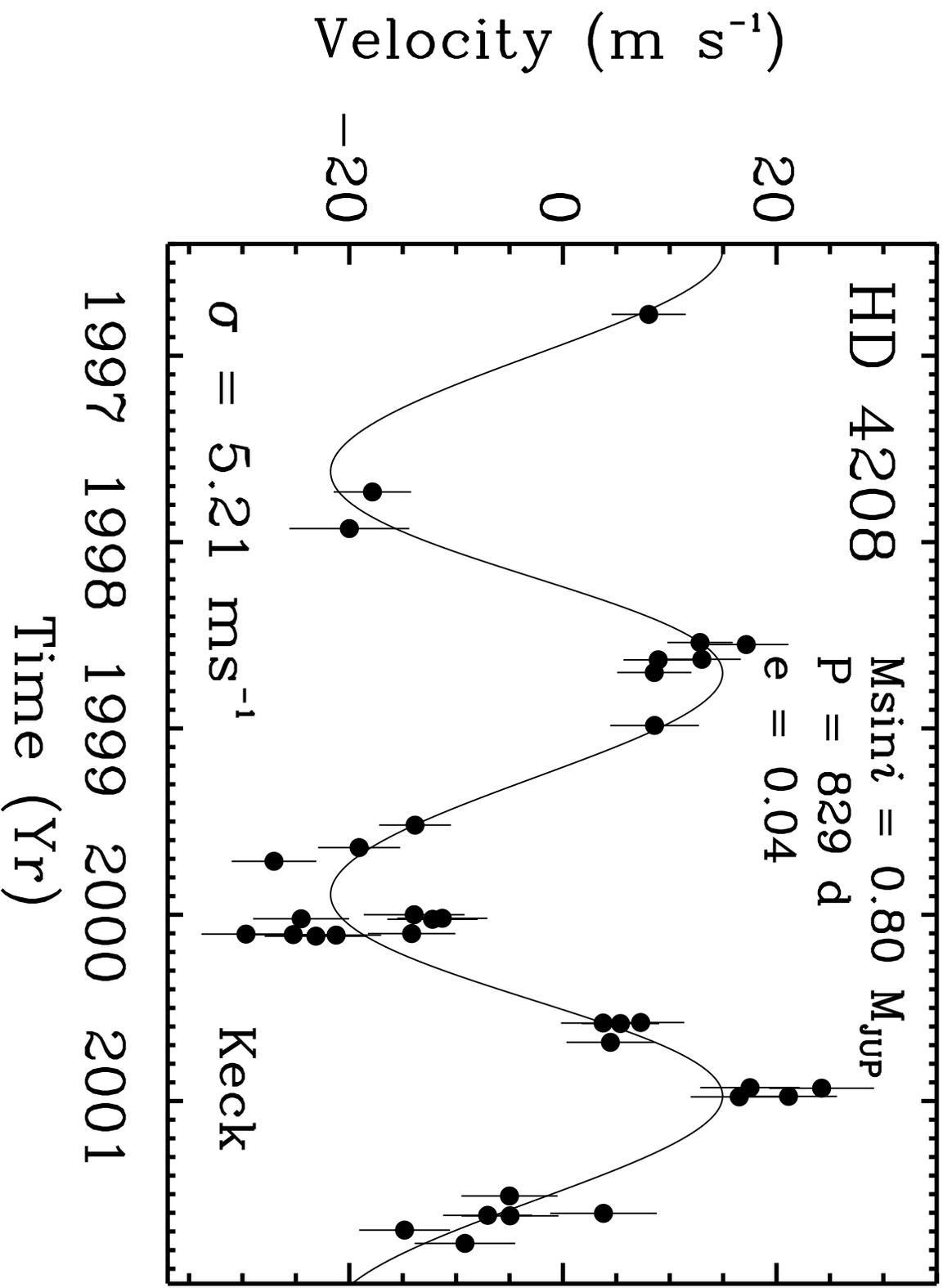
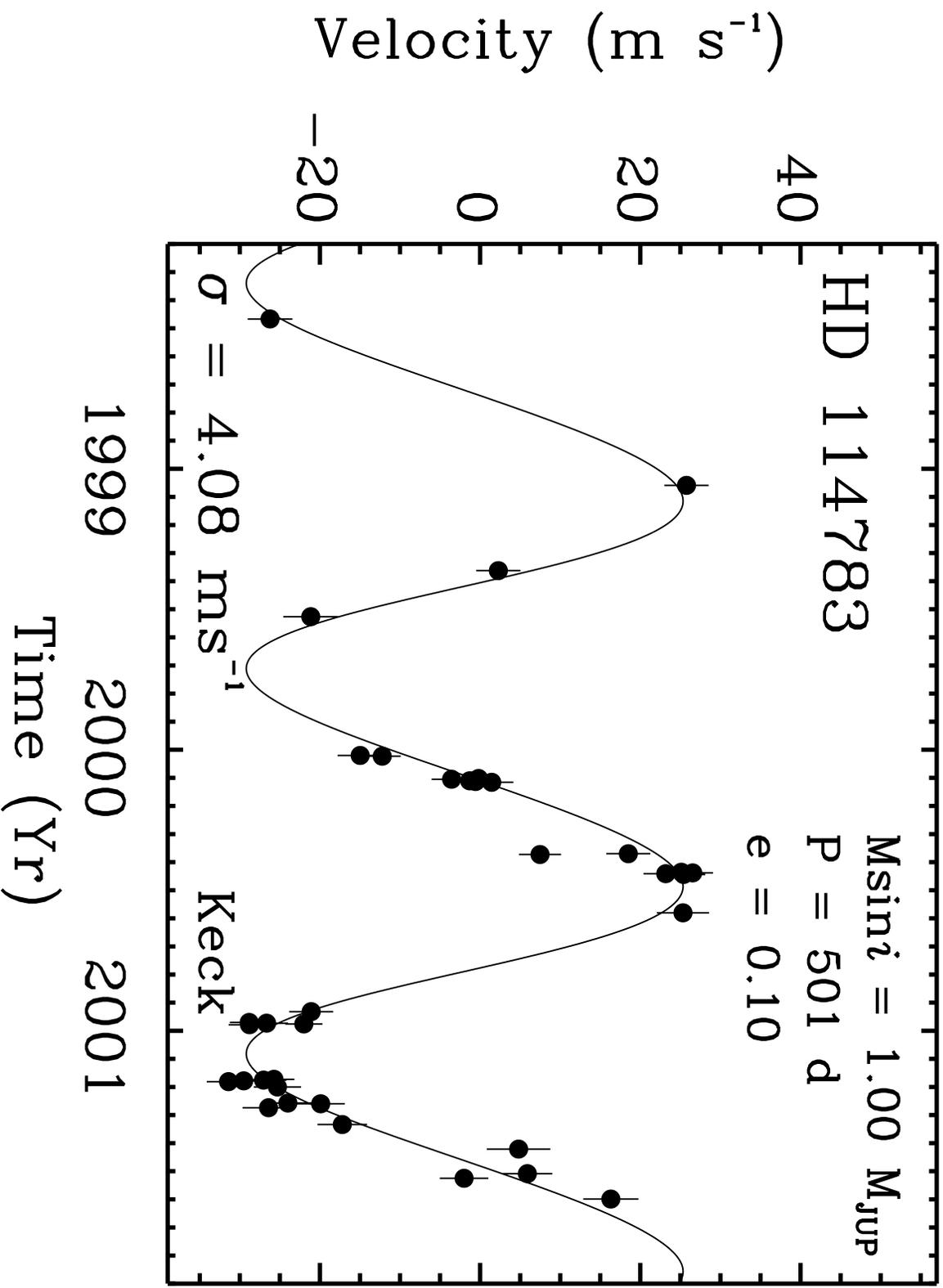
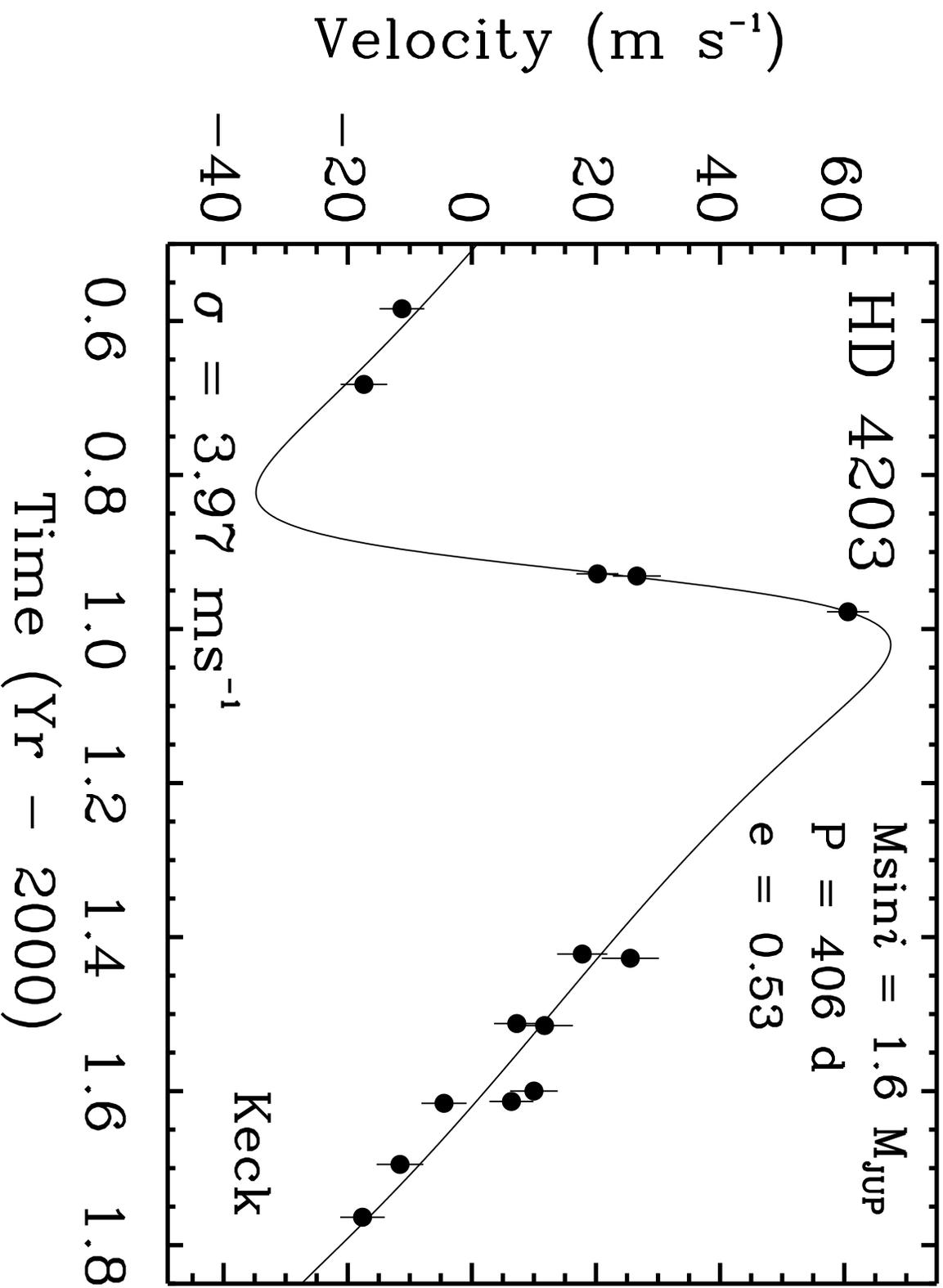
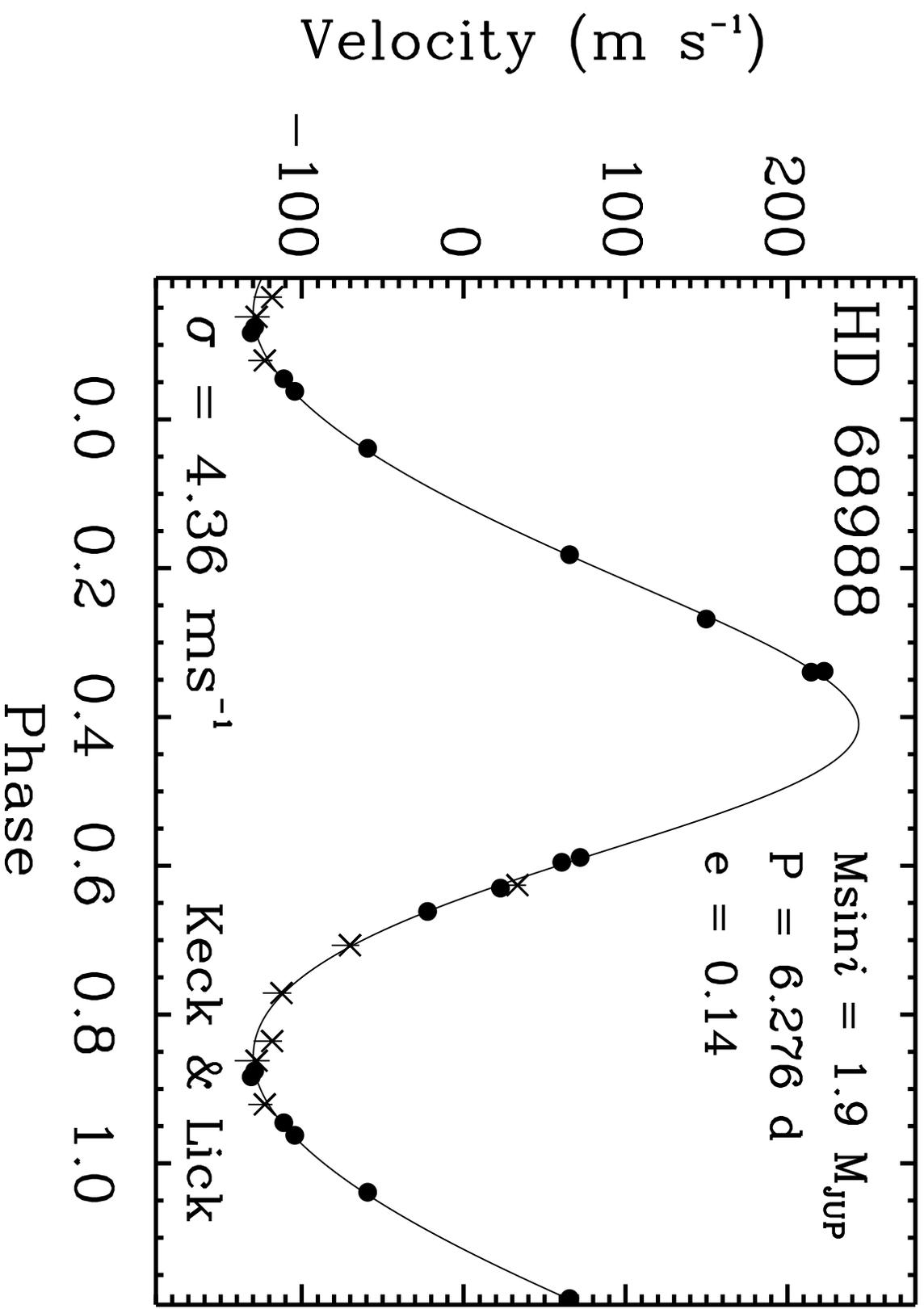
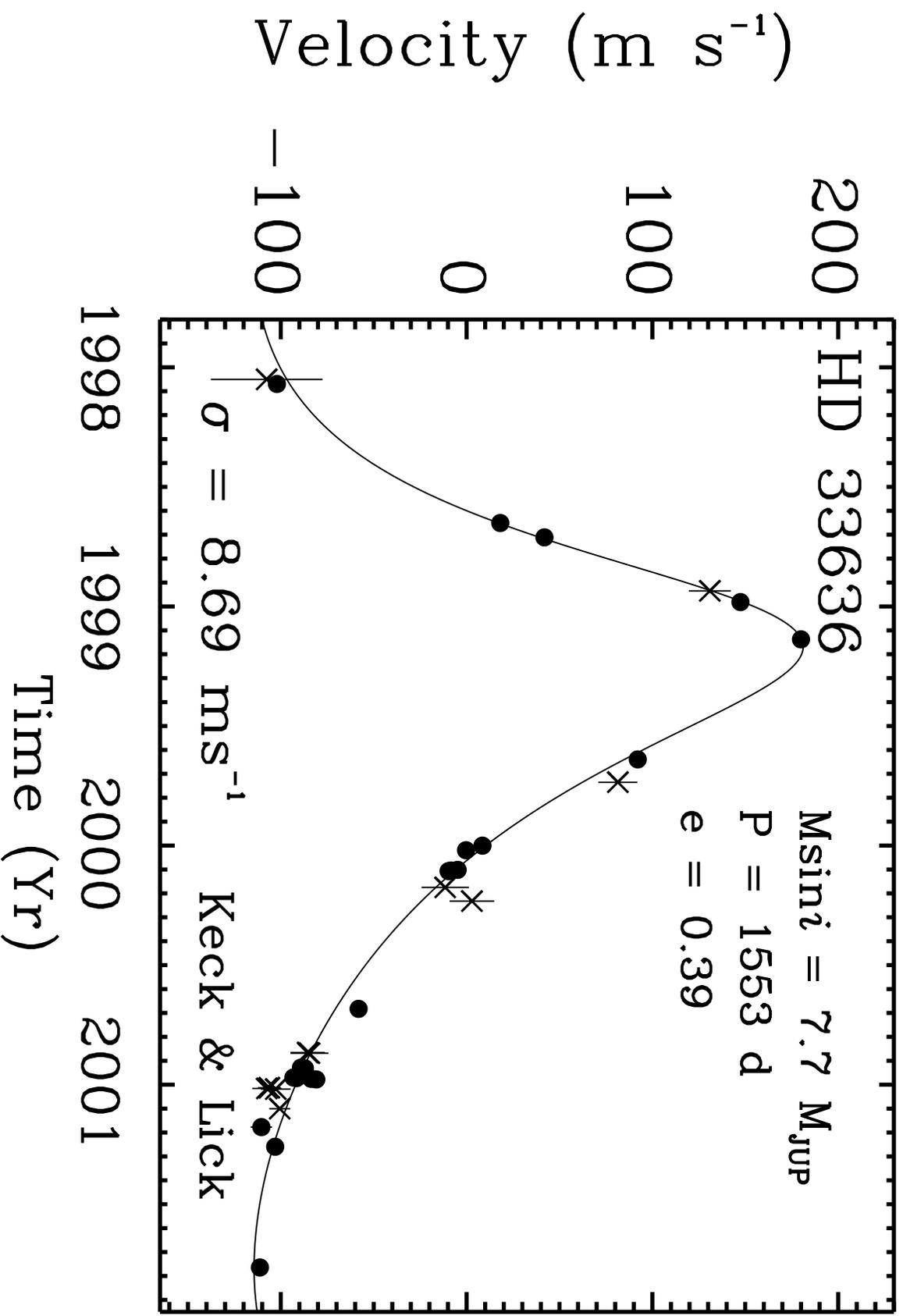
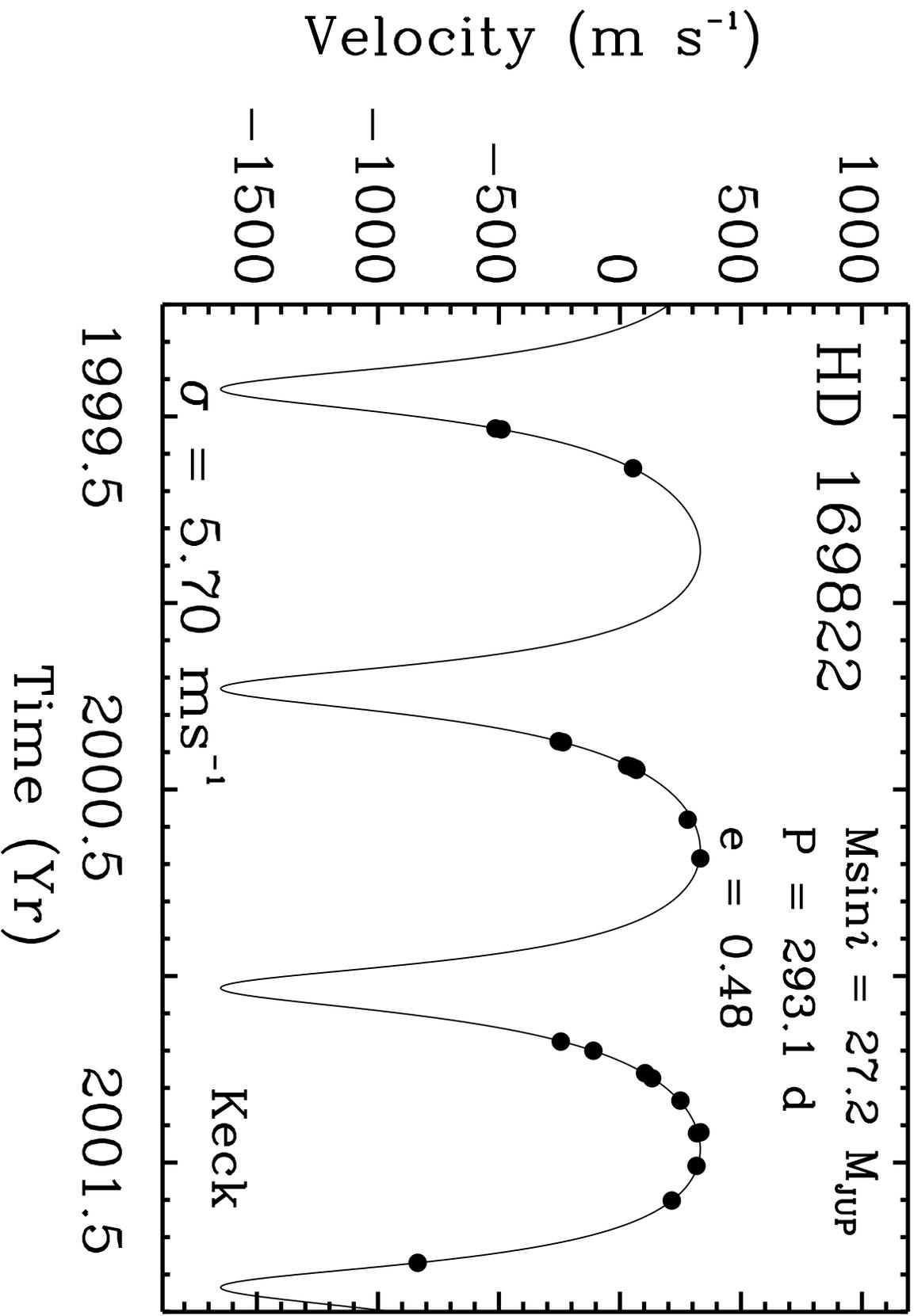
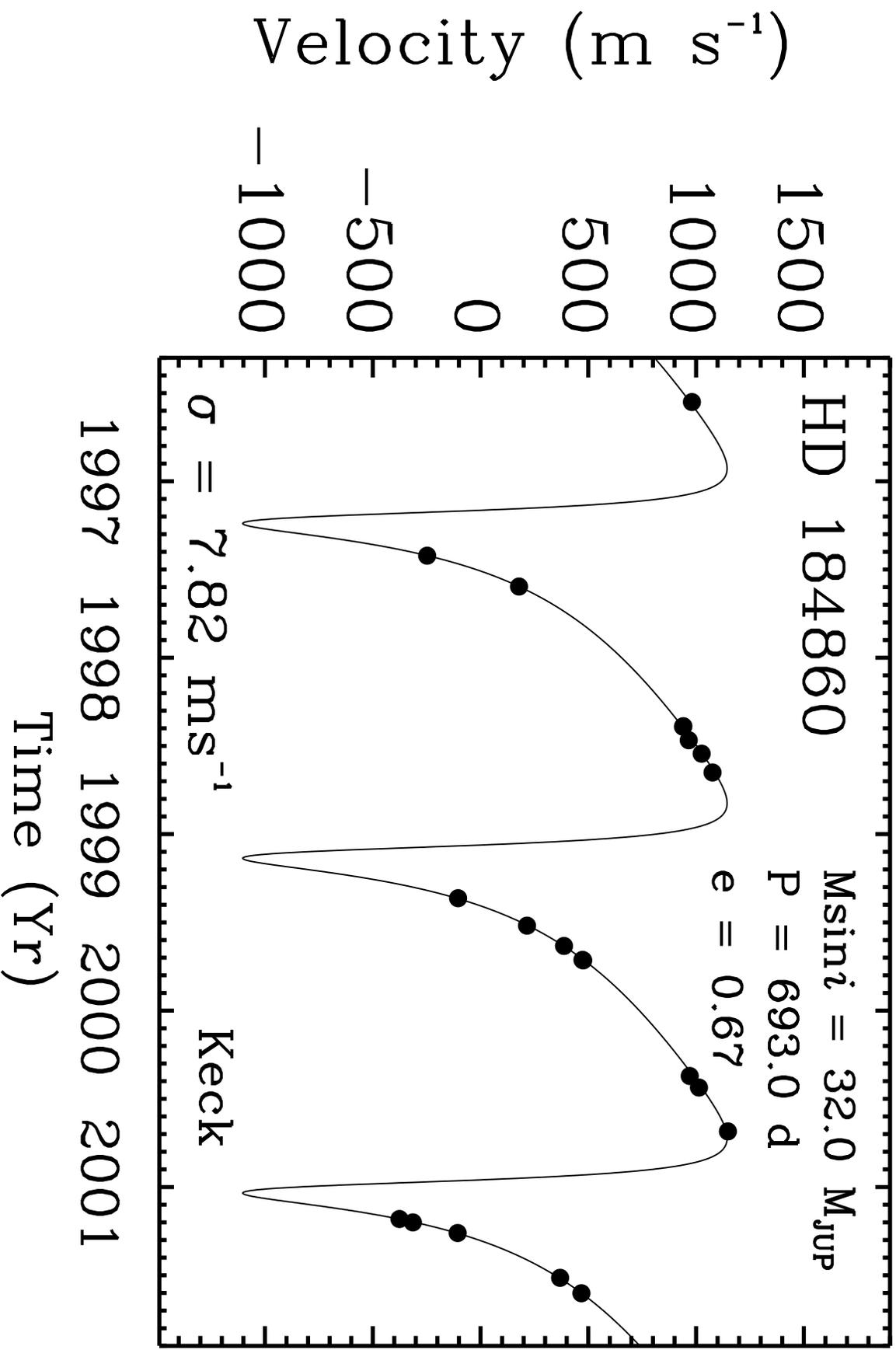
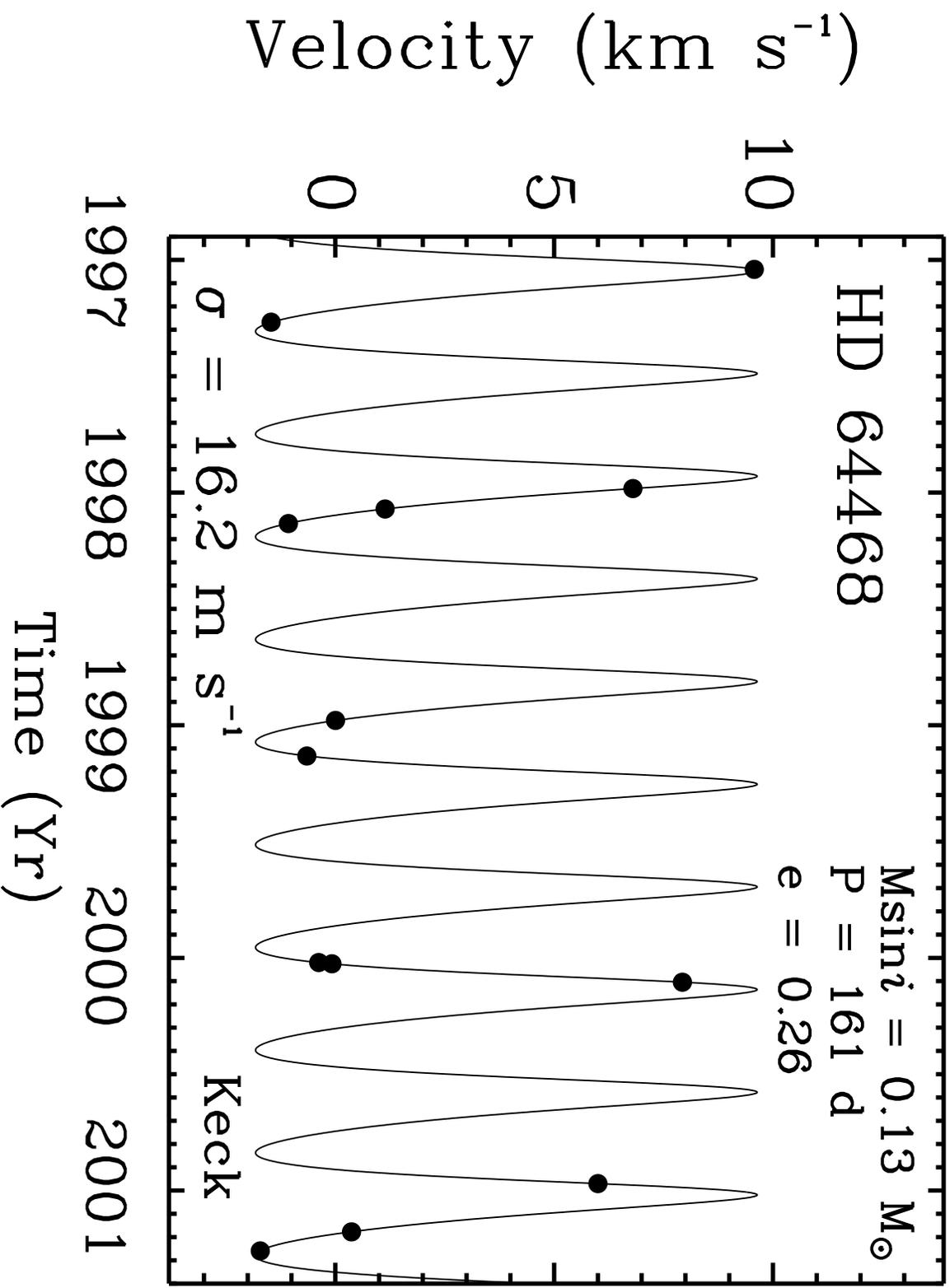
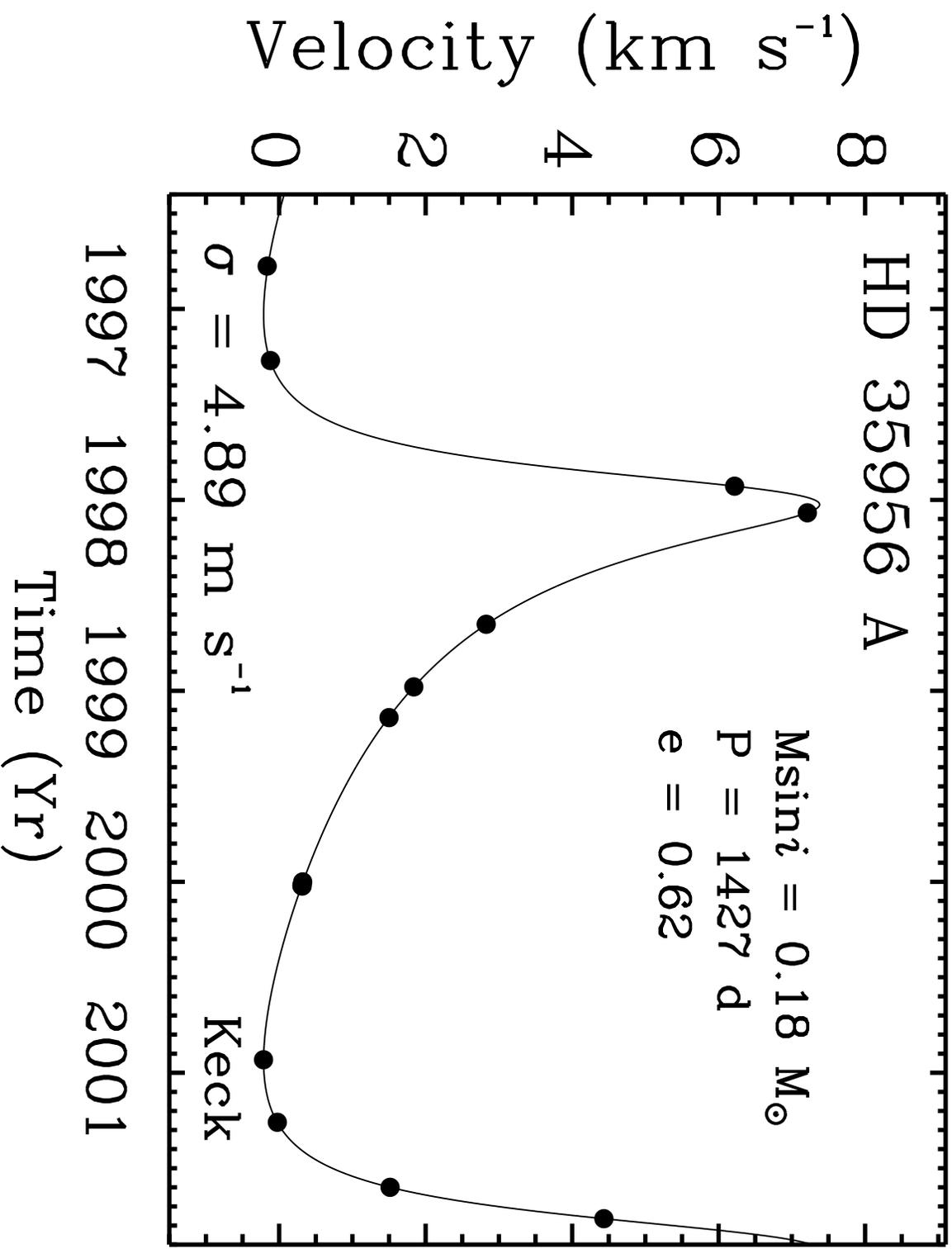
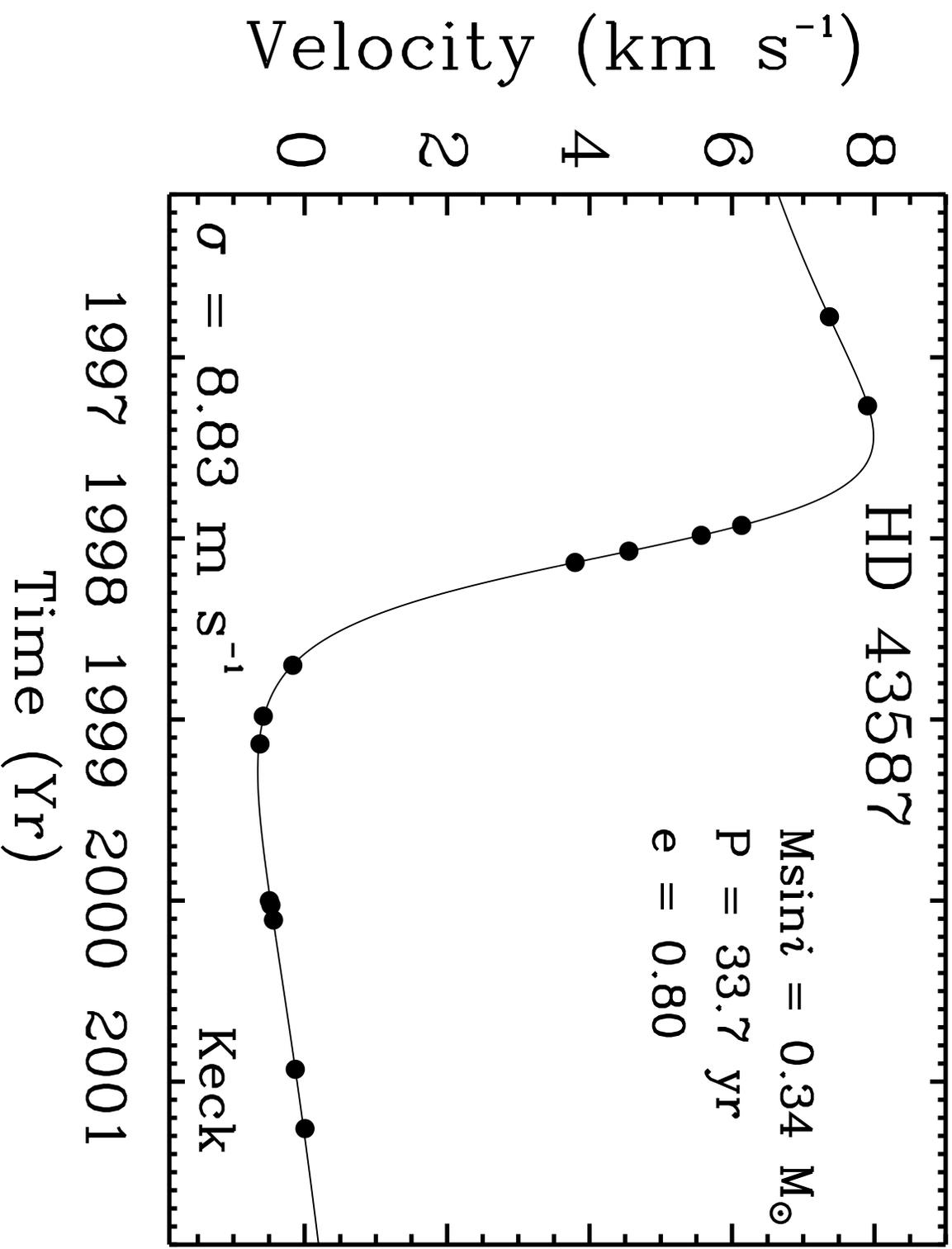
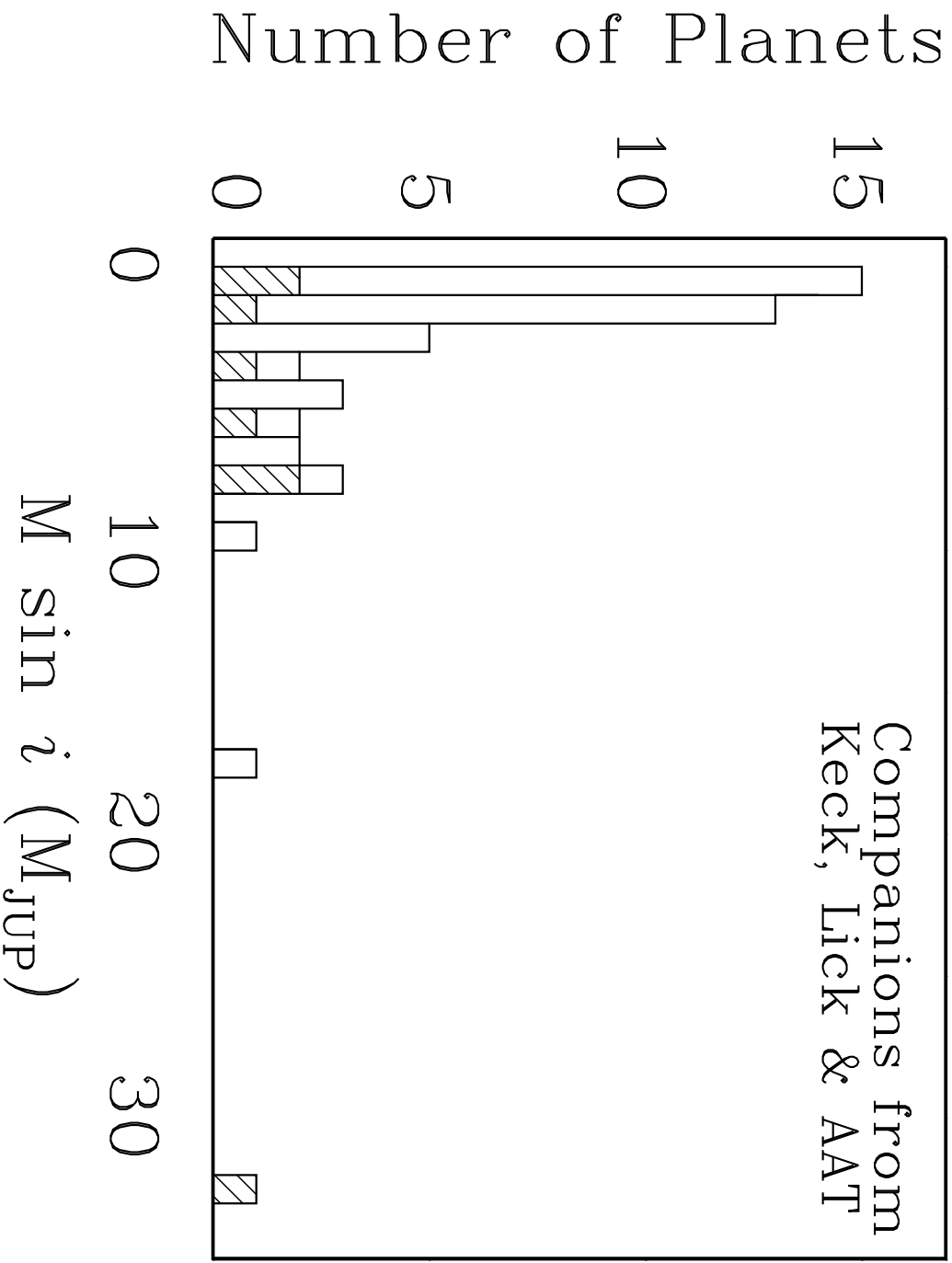
\end{document}